\documentstyle[preprint,eqsecnum,aps]{revtex}

\newcommand{\C}{\mbox{$\,${\sf I}\hspace{-1.2ex}{\bf C}}}
\newcommand{\D}{\mbox{$\,$/\hspace{-1.4ex}D}}
\begin{document}
\def\be{\begin{equation}}
\def\ee{\end{equation}}
\def\mn{\mu\nu}
\def\astf{{}^*f_{\mn}}
\def\orD{\overrightarrow{D}_\mu}
\def\olD{\overleftarrow{D}_\mu}
\def\ore{\overrightarrow{e}}
\def\orv{\overrightarrow{v}}
\def\orh{\overrightarrow{h}}
\def\ora{\overrightarrow{a}}
\def\B{{\cal B}}
\def\F{{\cal F}}
\def\P{{\cal P}}
\def\L{{\cal L}}
\def\H{{\cal H}}
\def\W{{\cal W}}
\def\M{{\cal M}}
\def\R{{\cal R}}
\def\pa{\partial}
\def\tn{\tilde\nabla}
\def\a{\alpha}
\def\d{\delta}
\def\l{\lambda}
\def\r{\rho}
\def\ga{\gamma}
\def\om{\omega}
\def\ka{\kappa}
\def\si{\sigma}
\def\w{\mbox{w}}
\def\vphi{\varphi}
\def\te{\tilde\eta}
\def\ha{\frac{1}{2}}
\def\ve{\varepsilon}
\def\px{\psi(x)}
\def\P{{\cal P}}
\def\H{{\cal H}}
\def\hpss{{\hps}_{s_3}}
\def\Us{U^{(s)}}
\def\Vpm{V^\pm_m}
\def\Mpm{{\cal M}^\pm_m}
\def\tohoch^#1{\mathrel{\mathop{\longrightarrow}\limits^{#1}}}
\def\ua{\underline{a}}
 \def\Gam{\Gamma_\mu}
\draft
\preprint{MPI-PhT/97-87}
\title{Broken Weyl-Invariance and the Origin of Mass}
\author{W. Drechsler\footnote{wdd@mppmu.mpg.de} and H. Tann}
\address{Max-Planck-Institut f\"ur Physik\\
         F\"ohringer Ring 6, 80805 M\"unchen, Germany}
\maketitle
\begin{abstract}
\renewcommand{\baselinestretch}{1.0}
\small\normalsize
A massless Weyl-invariant dynamics of a scalar, a Dirac spinor, and
electromagnetic fields is formulated in a Weyl space, $W_4$, allowing for
conformal rescalings of the metric and of all fields with nontrivial Weyl
weight together with the associated transformations of the Weyl vector
fields $\ka_\mu$ representing the $D(1)$ gauge fields with $D(1)$
denoting the dilatation group.
To study the appearance of nonzero masses in the theory
the Weyl-symmetry is broken explicitly and the corresponding reduction
of the Weyl space $W_4$ to a pseudo-Riemannian space $V_4$ is investigated
assuming the breaking to be determined by an expression involving the
curvature scalar $R$ of the $W_4$ and the mass of the scalar, 
selfinteracting field. Thereby also the spinor field acquires a mass
proportional to the modulus $\Phi$ of the scalar field in a Higgs-type
mechanism formulated here in a Weyl-geometric setting with $\Phi$
providing a potential for the Weyl vector fields $\ka_\mu$. After the
Weyl-symmetry breaking one obtains generally covariant and $U(1)$
gauge covariant field equations coupled to the metric of the
underlying $V_4$. This metric is determined
by Einstein's equations, with a gravitational coupling constant
depending on $\Phi$, coupled to the energy-momentum tensors
of the now massive fields involved together with the
(massless) radiation fields.
\end{abstract}
\newpage
\section{Introduction}
In order to understand the origin of nonzero masses in physics it is
suggestive to start from a theory, having no intrinsic scales,
which is formulated in a Weyl space, i.e.\ in a space-time $W_4$ being
equivalent to a family of Riemannian spaces
\begin{equation}
(g_{\mu\nu},\ka_\r),~(g'_{\mu\nu},
\kappa'_\r),~(g''_{\mu\nu},\kappa''_\r)\ldots
 \label{11}
\end{equation}
characterized by a class of metric tensors, $g_{\mu\nu}(x)$, and
Weyl vector fields, $\kappa_\r (x)$, related
by
\begin{eqnarray}
g'_{\mu\nu}(x)&=&\si(x)\, g_{\mu\nu} (x)\,,\\ \label{12a}
\kappa'_\r(x)&=&\kappa_\r(x) +\pa_\r \log\si (x)\,,\label{12b}
\end{eqnarray}
where $\si (x)\!\in\!D(1)$, with the dilatation group $D(1)$
being isomorphic to $R^+$ (the positive real line). The transformations
(1.2) and (1.3) are called {\sl Weyl transformations} involving
a conformal rescaling of the metric given by (1.2) together
with the transformation (1.3) of the Weyl vector field. A Weyl
space is thus characterized by {\sl two} fundamental forms:
\begin{equation}
ds^2= g_{\mu\nu}(x) dx^\mu\otimes dx^\nu\/;\qquad
\/\ka =\ka_\mu (x) dx^\mu\,,
\label{13}
\end{equation}
with the quadratic differential form $ds^2$ describing the distances on the
space-time manifold and with the linear differential form $\ka$ determining
the local change of the unit of length used for
measuring the distances. (Compare Weyl's original proposal of 1918
\cite{W} as well as his book {\sl Raum-Zeit-Materie} \cite{RZM}.
For a general discussion of Weyl spaces and conformal symmetry
in physics see Fulton, Rohrlich and Witten \cite{FRW}. Of course, we
shall here not identify the Weyl vector fields $\ka_\mu$ with the
electromagnetic potentials $A_\mu$ as Weyl originally did.)
In the following
we shall use a Weyl space of dimension $d=4$ with Lorentzian signature
$(+,-,-,-)$ of its metric. We collect the relevant formulae
characterizing the geometry of a Weyl space in Appendix A -- C.

A $W_4$ reduces to a pseudo-Riemannian space $V_4$ for
$\ka_\mu=0$; a $W_4$ is equivalent to a $V_4$ if the
``length curvature" (Weyl's ``Streckenkr\"ummung'') associated
with the Weyl vector fields $\ka_\mu$ is zero, i.e.
\begin{equation}
f=d\ka=0\,, \label{14}
\end{equation}
and the transfer of the unit of length is integrable.

We shall use in this paper a Weyl geometry as a geometric stratum
for the formulation of a physical theory involving, at the beginning,
only massless fields. The theory presented in Sect.\ II is characterized
by a Weyl-invariant Lagrangean density, $\L_{W_4}$, involving
besides the fields $g_{\mn}$ and $\ka_\r$ a complex massless scalar
field $\vphi$ as well as a massless Dirac spinor field $\psi$. The local
group structure of the theory is given by the gauge group
$SO(3,1)\otimes D(1)$ --~ or rather by $Spin(3,1)\otimes D(1)$ for the
Dirac spinors. (Compare the definition of the relevant frame bundles
given in Appendix A and in Sect.\ II below.)
In Sect.\ III we extend the description by including electromagnetism
$(A=A_\mu (x) dx^\mu$; $F=dA$) using the minimal electromagnetic
coupling procedure yielding a Lagrangean density, $\tilde{\L}_{W_4}$,
for a massless theory with gauge group $SO(3,1)\otimes D(1)\otimes U(1) \simeq
SO(3,1)\otimes Gl(1,\C)$, or its covering group $Spin(3,1)\otimes D(1)
\otimes U(1)$ for the Dirac spinor fields. In Sect.\ IV this $U(1)$
and $D(1)$ gauge covariant as well as generally
covariant -- i.e.\ Weyl-covariant -- theory involving ``gravitation''
which describes a Weyl-symmetric
dynamics for the metric in the form of field equations of 
Einstein type,
is broken to a theory defined on a space-time $V_4$ by the introduction of a
Weyl-symmetry breaking term depending on the 
scalar curvature invariant of the
$W_4$ giving thereby a nonzero mass to the scalar field $\vphi$.
This reduction $W_4\to V_4$ is governed by a Weyl vector field of the
type $\ka_\r=\pa_\r \chi $ being ``pure gauge'', i.e.\ being
given by a gradient of a scalar function yielding thus zero length
curvature: $f=0$. At the same
time also the Dirac field acquires a mass due to a Yukawa-type coupling
between the $\vphi$ and the $\psi$ fields allowed by Weyl-invariance and
the adopted Weyl weights of the fields involved (see below). An
essential part in the Weyl-symmetry breaking and the concomitant
mass generation is the requirement that Einstein's equations for the
metric are obtained in the limiting $V_4$ description in coupling
ultimately the purely metric geometry to the energy-momentum
tensors of the massive fields $\vphi$ and $\psi$ as well as to the
radiation field F. Our aim in this paper is thus to study a
Higgs-type phenomenon in a geometric setting containing gravitation
from the outset -- or, more specifically -- to investigate such a
mass giving phenomenon as the breaking of a Weyl-symmetry yielding a
Riemannian description in the limit.

The work presented in this paper is an extension of an
investigation of a theory involving only a complex scalar quantum
field, $\vphi$, which was carried out by one of us \cite{HT}.
It turned out in this investigation that the $\vphi$-field did not
behave as a bona fide matter field in this Weyl geometric framework.
When split into a modulus $\Phi (x)$ and a phase $S(x)$, i.e.
\begin{equation}
\vphi =\Phi e^{{i\over\hbar}S}\/; ~~\/\Phi=\sqrt{\Phi^2}\mbox{~ with~ }
\Phi^2=\vphi^*\vphi\/, \label{15}
\end{equation}
the $\Phi$- and the $S$-part of $\vphi$ played different roles
in this theory with $\Phi$ being affected by the $D(1)$ (i.e.\ Weyl)
gauge degree of freedom related to the $W_4$
geometry while $S$ being, as
usual, affected by the $U(1)$ (electromagnetic) gauge degree of freedom.
Besides the fact that the energy-momentum tensor for the $\vphi$-field
automatically appears in the modified ``new improved'' form
(compare Callan, Coleman and Jackiw \cite{CCJ}) in the adopted
$W_4$ framework the modulus field $\Phi$ could be related to the
Weyl vector field $\ka_\r$ and -- in this way -- played a geometric
role in the theory. In the present paper we want to study the symmetry
breaking yielding a $V_4$ from a $W_4$ by including an additional
Dirac spinor field $\psi$ into the theory, which is considered as a genuine
matter field, and determine the mass the $\psi$-field acquires in the
symmetry reduction involving only a mass giving term for the
$\vphi$-field in connection with a curvature invariant of the
underlying Weyl geometry (the Weyl curvature scalar, $R$, as
already mentioned above). To study the implications of the presence
of a spinor field in this Weyl geometric setting
is the main theme of the present paper.

However, we do not consider in this context the case of a Weyl-geometric
description of spinors alone, nor do we touch upon the question whether
there exist spinor geometries of Weyl type which are not related to
Riemannian geometry \cite{T}. Since we use the Weyl-geometric
framework as a mathematical tool to construct, ultimately, a physical
theory with nonzero masses from a Weyl-symmetric massless theory
for bosonic {\sl and} fermionic fields,
we are only interested in those aspects of a Weyl geometry involving
spinor fields which naturally connect to a pseudo-Riemannian theory
with nonzero masses for the spinor fields. Hence the Weyl-symmetry
describes a massless scenario behind 
the curtain of our real world which itself
is characterized by a broken Weyl-symmetry with massive source terms
representing ponderable matter generating gravitation as implied by
Einstein's general theory of relativity. We also include electromagnetism
in this description (Sect.\ III) and the presence of massless
radiation fields before and after Weyl-symmetry breaking. 

The theory we present
in this paper is certainly not fully realistic. In order to make
closer contact with the Higgs phenomenon of the standard model
in particle physics it would be necessary to introduce an
additional local weak isospin group $SU(2)$ into the Weyl framework
as well as a corresponding representation character for
the scalar field and for the chiral
fermion fields possessing particular hypercharges. We shall not do this
in this paper in order to study the geometric mass giving phenomenon
in its simplest and most transparent form for electrically charged
scalar and spinor fields and determine the role played by the Weyl
vector field $\ka_\r$ in this context. It appears that the modulus
$\Phi$ of the scalar field $\vphi$ acts like a Higgs field
determining $\ka_\r$ in the limiting $V_4$ theory with vanishing
length curvature, i.e.\ with $f=0$. This is concluded from a
consideration of the trace and divergence relations following from
the Weyl-covariant equations for the metric in the unbroken
theory with a Yukawa-type coupling between the $\vphi$ and $\psi$
fields. Moreover, an essential part in this discussion is the use of
the contracted Bianchi identities for a
$W_4$ as well as the square of the Dirac operator. This
investigation is carried out in Sect.\ II and, including
electromagnetism, in Sect.\ III. In its broken form the divergence
relations -- with the physical meaning of energy and momentum balance
equations -- are, finally, studied in Sect.\ IV. The determination of
the free parameters of the theory and the relation to the field
equations of general relativity are given in Sect.\ V. We end the
discussion with some final remarks on the results obtained which
are presented in Sect.\ VI.
\section{Weyl-Invariant Lagrange Theory}

It is well-known that the scalar wave equation in a $V_4$ for a massless
field $\vphi$ reading
\begin{equation}
\Box \vphi +{1\over 6} \bar R\vphi =0
\label{21}
\end{equation}
is conformally invariant \cite{RP}, i.e.\ it is
invariant against conformal rescalings of the metric of the
underlying Riemannian space-time according to Eq. (1.2) provided
the field $\vphi$ has the conformal weight $w(\vphi)=-\ha$ (in
a four-dimensional space).
In (\ref{21}) the scalar curvature of the $V_4$ is denoted by
$\bar R$, and $\Box =g^{\mn}\bar\nabla_\mu\pa_\nu$
is the d'Alembert operator. (For the notation used in this and
the subsequent sections see Appendix A.)

We want to study in this paper conformal rescalings of the metric in
a Weyl-geometric setting, i.e.\ as part of the Weyl-symmetry expressed
by Eqs.~(1.2) and (1.3) involving the metric $g_{\mn}$
{\sl and} the Weyl vector field $\ka_\r$. We thus want to extend the
field equation for a complex massless scalar field $\vphi$ to a Weyl space $W_4$.
At the same time we include a massless Dirac spinor field $\psi$ as a
representative of fermionic matter and investigate a
Weyl-invariant dynamics of the four fields $g_{\mn},\ka_\r,\vphi$ and
$\psi$. Here the first two fields represent the geometric side of the
problem while $\vphi$ and $\psi$ are, so to speak, bosonic and fermionic
degrees of freedom representing the matter side of the dynamics.
However, this division into geometric and material aspects is superficial in a
massless Weyl-invariant formulation since all fields having
particular nonzero Weyl weights transform nontrivially under Weyl
transformations and thus the geometric fields mix with the massless
$\vphi$ and $\psi$ fields.

The Weyl weights of $g_{\mn}$ and $\ka_\r$ are 1 and 0, respectively (see
Appendix A). The Weyl weight of $\vphi$ is taken to be $w(\vphi)=-\ha$,
as mentioned above, and the Weyl weight of the Dirac field $\psi$ is chosen
to be $w(\psi)=-{3\over 4}$ (compare Pauli \cite{P} in this context).
This last choice is dictated by the fact
that the Dirac vector current, $j^{(\psi)}_\mu$, (see Sect.\ III below)
should have Weyl weight $-1$ so that this current can act as a source
current in Maxwell's equations. (The coupling of the above described
dynamical system to the electromagnetic fields yielding a complete
massless dynamics including electromagnetic interactions and
radiation fields $F_{\mn}$,
described by a Weyl- {\it and} $U(1)$ covariant
gauge theory including gravitation,
will be studied in detail in Sect.\ III). Ultimately, the Weyl-invariance
has to be broken since there are nonzero masses and corresponding length
scales appearing in nature. However, our proposal is to approach the
realistic physical world from a massless Weyl invariant description in
order to investigate, from a geometric point of view, how a mass or length scale
may be established in a physical theory by the breaking of a symmetry
characterizing a massless situation. At the same time the capacity
of matter and electromagnetic radiation of generating gravitational
fields will be included in the theory as an essential structural
element leading, ultimately, to Einstein's equations in the broken 
version of the theory.

Although our motivation in this endeavour is, as mentioned, geometric
in its origin we shall use the Lagrangean method -- formulated against
the adopted Weyl geometric background -- in order to specify a particular
theory: The Lagrangean density $\L_{W_4}$ should be a scalar of Weyl
weight zero depending on $g_{\mn},\ka_\r,\vphi$, and $\psi$ and their
first derivatives to yield a set of second-order Weyl-covariant field equations
characterizing the massless dynamics. We shall see that the postulated
Weyl-invariance of $\L_{W_4}$ and the construction of $\L_{W_4}$ in
terms of Weyl-covariant derivatives of the quantities involved (see
Appendix A) limits the number of possibilities considerably.

The Weyl-invariant Lagrangean on which we shall base our subsequent
discussion in the massless case, neglecting electromagnetism, is the
following scalar hermitean density of Weyl weight zero:
\begin{eqnarray}
{\cal L}_{W_4}= &&K\sqrt{-g}\biggl\{  \ha g^{\mn}D_\mu\vphi^*
D_\nu\vphi -{1\over {12}}
R\,\vphi^*\vphi +\tilde\a R^2-\beta (\vphi^*\vphi)^2+
\nonumber\\ &&\qquad +{i\over 2}
(\bar\psi \ga^\mu\overrightarrow{D}_\mu\psi -\bar\psi\overleftarrow{D}_\mu
\ga^\mu\psi)+\tilde\ga\sqrt{\vphi^*\vphi}(\bar\psi\psi) -
\tilde\delta {1\over 4}f_{\mn}f^{\mn}\biggr\}\/,\label{22}
\end{eqnarray}
with
\begin{equation}
D_\nu\vphi=\pa_\nu\vphi +\ha \ka_\nu\vphi\/~,\label{23}
\end{equation}
and
\begin{eqnarray}
\orD\psi &=& \overrightarrow{\nabla}_\mu\psi+{3\over 4}\ka_\mu\psi =
\left(\overrightarrow{\pa_\mu}+i\Gamma_\mu +
{3\over 4}\ka_\mu \right)\psi\,,\\ \label{24}
\bar\psi\olD &=& \bar\psi\overleftarrow{\nabla}_\mu+
{3\over 4}\ka_\mu\bar\psi =
\bar\psi \left(\overleftarrow{\pa_\mu} -i\Gamma_\mu+
{3\over 4}\ka_\mu \right)\/~
\label{25}
\end{eqnarray}
being the Weyl-covariant derivatives of the scalar and spinor fields,
respectively, where $\nabla_\mu$ denotes the covariant derivative with
respect to the $W_4$-connection with coefficients defined in (A1).
Moreover
\begin{equation}
\Gamma_\mu=\Gam (x)=\lambda^j_\mu (x)\ha\Gamma_{jik} (x) S^{ik}\/,
\label{26}
\end{equation}
\begin{equation}
\mbox{with~~ } S^{ik}={i\over 4}[\ga^i,\ga^k]\/~\label{27}
\end{equation}
is the spin connection [where the local Lorentz index $j$
is turned into a Greek index $\mu$ with the help of the vierbein
fields $\lambda^j_\mu (x)$], i.e.\ is the connection on the
associated  spinor bundle
\begin{equation}
S_{\w}=S_{\w}\Bigl(W_4,\C_4, Spin(3,1)\otimes D(1)\Bigr)
\,,\label{28}
\end{equation}
with fiber $F=\C_4$ being a representation space for the
Dirac spinors and with the structural group $Spin (3,1)\otimes D(1)$.
A spinor field $\psi (x)$ is a section of $S_{\w}$ with the usual local
action of the group $Spin(3,1)$, and with $D(1)$ acting on the absolute
value of $\psi (x)$, i.e. on the scalar $\sqrt{\bar\psi(x)\psi(x)}$
with Weyl weight $-{3\over 4}$ [compare (A2)].

The bundle $S_{\w}$ is associated to the spin frame bundle over $W_4$
\begin{equation}
\bar P_{\w}=\bar P_{\w}\Bigl(W_4,\bar G=Spin(3,1)\otimes D(1)\Bigr)
\,, \label{29}
\end{equation}
which is related to the Weyl frame bundle defined in Eq.\ (A6) in the
usual way by lifting the homomorphism between the universal covering
group $Spin(3,1)$~and the orthochronous Lorentz group $SO(3,1)$
to the bundles $\bar P_{\w}$ and $P_{\w}$. The homomorphism between the
structural groups of the two bundles (with the $D(1)$ factor being
untouched) is provided by the well-known formula
\begin{equation}
S(\Lambda)\ga^iS^{-1}(\Lambda)=
[\Lambda^{-1}]^i{}_k\ga^k\/,\label{210}
\end{equation}
where $S(\Lambda)\in\! Spin(3,1)$, and
$S^{-1}(\Lambda)=\ga^0 S^\dagger(\Lambda)\ga^0$
with $S^\dagger(\Lambda)$ denoting the adjoint of $S(\Lambda)$, and $\Lambda\in
SO(3,1)$. The $\Gamma_{jik}$ in Eq.\ (\ref{26}) are the connection
coefficients defined in Eqs.\ (A8) and (A10), and the $\ga^i;i=0,1,2,3$ are the
constant Dirac matrices obeying
\begin{equation}
\{\ga^i,\ga^k\}=\ga^i\ga^k+\ga^k\ga^i=
2\eta^{ik}{\bf 1}\,.\label{211}
\end{equation}
The $\ga$-matrices with upper Greek indices appearing
in (\ref{22}) are $x$-dependent
quantities of Weyl weight $-\ha$ defined by
\begin{equation}
\ga^\mu =\ga^\mu (x)=\l^\mu_i(x)\ga^i\mbox{~~ obeying~~ }
\{\ga^\mu,\ga^\nu\}=2 g^{\mn}{\bf 1}\,.
\label{212}
\end{equation}
Finally, $R$ in (\ref{22}) denotes the scalar curvature of the $W_4$
having Weyl weight $w(R)=-1$, which is defined in (A31). All the terms
in the curly brackets of Eq.~(\ref{22}) have Weyl weight $-2$ which,
together with $w(\sqrt{-g})=+2$, yields Weyl weight zero for $\L_{W_4}$.

The first term in the curly brackets of (\ref{22}) is the kinetic part
for the complex $\vphi$ field. The second term is the contribution
guaranteeing conformal invariance in a $V_4$ limit. The third term with
constant $\tilde\alpha$ (of dimension [$L^2$]; $L$=length)
is a term included to yield a nontrivial dynamics for the $\ka_\r$-fields.
The fourth term multiplied by a constant $\beta$ (of dimension
[$L^{-2}$]) is the nonlinear $\vphi^4$-coupling of the $\vphi$-field
allowed by the Weyl weight $w(\vphi )=-\ha$.
The fifth term is the kinetic term for the Dirac field. The sixth term
is a Yukawa-like coupling of $|\vphi|=\sqrt{\Phi^2}$ to ($\bar\psi\psi$)
with constant $\tilde\gamma$ (of dimension [$L^{-1}$]) \footnote{One
could also think of writing the Yukawa-like coupling term of Weyl weight
$-2$ between the $\vphi$- and the $\psi$-fields in the form
$\tilde\gamma'\ha (\vphi +\vphi^*)(\bar\psi\psi )$ yielding
qualitatively the same results as the $\tilde\ga$-coupling used in
the text which, however, has the advantage of being independent
of the phase $S$ of $\vphi$.}.
Finally, the last term, multiplied by a constant $\tilde\delta$ (of
dimension [$L^2$]), is the contribution of the $f$-curvature, i.e.
of the free $\ka_\r$ fields to the total
Lagrangean. [We remark in parenthesis that the quadratic curvature
invariants of Weyl weight zero for a $W_4$ are investigated at the
end of Appendix A and in Appendix C. It appears from this discussion
(compare Eq. (A54)) that the invariant $\sqrt{-g}f_{\mn}f^{\mn}$ plays
a special role compared to the other quadratic curvature invariants
for a $W_4$. For this reason we include in (\ref{22}) besides the
term $\sim\! \sqrt{-g}R^2$ only the term
$\sim\!\sqrt{-g}f_{\mn}f^{\mn}$ in close analogy
to electromagnetism.]
The length dimension of the field $\vphi$ is here
assumed to be zero. Relative to this the fermion field has length
dimension $[L^{-\ha}]$. The overall constant $K$ (with
dimension [energy $\cdot L^{-1}$]) is a factor converting the
length dimension of the expressions in the curly brackets (which is
$[L^{-2}]$) into [energy $\cdot L^{-3}$] to give to $\L_{W_4}$ --
later after symmetry breaking -- the correct dimension of an energy/vol.
The $K$ factor drops out of the field equations for the Weyl-symmetric
case discussed in this section and appears in (\ref{22}) only
for convenience.\\

We now vary the fields in Eq. (\ref{22}) according to the usual
rules and determine the field equations from the variational
principle $\delta\!\int\!\L_{W_4}d^4x=0$. Using the notation for the
modulus of $\vphi$ introduced in Eq. (\ref{15}) the field equations in
Weyl-covariant form are:
\begin{eqnarray}
&&\delta\vphi^*:~~ g^{\mn}D_\mu D_\nu\vphi +{1\over 6} R\vphi + 4\beta
(\vphi^*\vphi)\vphi - \tilde\gamma {\vphi\over\sqrt{\Phi^2}}
(\bar\psi\psi)=0\/,\label{213}\\
&&\delta\psi^{\dagger} :~~ -i\gamma^\mu D_\mu\psi -\tilde\gamma\sqrt{\Phi^2}
\psi =0\/,\label{214}\\
&&\delta\ka_\r :~~ \tilde\delta D_\mu f^{\mu\rho}=-6\tilde\alpha D^\rho R\/,
\label{215}\\
&&\delta g^{\mn}:~~ {1\over 6}\Phi^2\left[R_{(\mn)}-\ha g_{\mn}R\right]
-4\tilde\alpha R\left[R_{(\mn)}-{1\over 4}g_{\mn} R\right]-4\tilde\alpha
\left\{D_{(\mu}D_{\nu)}R-g_{\mn}D^\r D_\r R\right\}=\nonumber\\
&&\quad\quad\qquad\qquad 
=\Theta^{(\vphi)}_{\mn}+T^{(\psi)}_{\mn}+T^{(f)}_{\mn}-g_{\mn}
\tilde\gamma\sqrt{\Phi^2} (\bar\psi\psi)\,,\label{216}
\end{eqnarray}
together with the complex conjugate equation of (\ref{213})
and the Dirac adjoint of (\ref{214}). Here we have used in the
last equation the following abbreviations for the symmetric
energy-momentum tensors of Weyl weight $-1$ for the $\vphi$-field,
the $\psi$-field, and the $f$-field, respectively, defined in a
Weyl-covariant manner:
\begin{eqnarray}
\Theta^{(\vphi)}_{\mn}&=&\ha (D_\mu\vphi^*D_\nu\vphi+D_\nu\vphi^*D_\mu\vphi )-
{1\over 6}\left\{D_{(\mu}D_{\nu)}\Phi^2-g_{\mn}D^\r D_\r\Phi^2\right\}
\nonumber \\
&&\qquad\qquad -g_{\mn}\left[\ha g^{\r\l}D_{\r}\vphi^*D_{\l}\vphi -
\beta (\vphi^*\vphi )^2\right]\,, \label{217}\\
T_{\mn}^{(\psi)}&=& {i\over 2}\left\{\bar\psi\gamma_{(\mu}
\overrightarrow D_{\nu)}
\psi -\bar\psi\overleftarrow D_{(\mu}
\ga_{\nu)}\psi\right\}-g_{\mn}{i\over 2}
\left\{\bar\psi\ga^\rho\overrightarrow D_\r
\psi-\bar\psi\overleftarrow D_\r\gamma^\r\psi\right\}\,,
\label{218}\\
T_{\mn}^{(f)} &=& -\tilde\delta \left[f_{\mu\si}f_\nu{}^\si -{1\over 4}
g_{\mn}f^{\rho\lambda}f_{\r\l}\right]\,.\label{219}
\end{eqnarray}

The $\vphi$- and $\psi$-equations (\ref{213}) and (\ref{214}) are
coupled via the $\tilde\ga$ term which lead to mass-like expressions in
both equations. We shall come back below to the status of these
coupling terms in the Weyl-symmetric framework (i.e. before symmetry
breaking) when we discuss the conditions following from Eq. (\ref{216}).
We remark in passing that the $\ka_\r$ field actually drops out of Eqs. (\ref{213})
and (\ref{214}) due to the identities
\begin{equation}
g^{\mn}D_\mu D_\nu\vphi +{1\over 6} R\phi= g^{\mn}
\bar\nabla_\mu\pa_\nu\vphi +{1\over 6}\bar R\vphi\/,\label{220}
\end{equation}
and
\begin{equation}
-i\ga^\mu D_\mu\psi = -i\ga^\mu\bar\nabla_\mu\psi\/,\label{221}
\end{equation}
which are valid for the Weyl weights chosen for $\vphi$ and $\psi$,
respectively. Here $\bar\nabla_\mu$ denotes the metric covariant
derivative [compare the discussion after Eq. (A1)]. Also on the lhs
of (\ref{215}) one can use the identity
\begin{equation}
D_\mu f^{\mu\r}=\bar\nabla_\mu f^{\mu\r}
\label{222}
\end{equation}
to eliminate the $\ka_\r$ contributions in the covariant derivative.
Eq.~(\ref{215}) represents a Maxwell-type equation coupling the field
strength $f^{\mu\r}$ to a current proportional to $D^\r R=\pa^\r R+
\ka^\r R$. It is easy to show by contracting (\ref{215}) with $D_\rho$,
i.e.\ in taking the Weyl-covariant divergence, that
\begin{equation}
\tilde\alpha D^\r D_\r R =0 \label{223}
\end{equation}
exhibiting the $W_4$-covariant current conservation following from
(\ref{215}) together with $\bar\nabla_\r\bar\nabla_\mu f^{\mu\r}=0$.

The equations (\ref{215}) and (\ref{216}) do contain $\ka_\r$ in a
complicated manner. They represent 4+10 coupled Weyl-covariant equations
for the determination of the Weyl vector field $\ka_\r$ and the metric
$g_{\mn}$ [modulo Weyl-transformations (1.2), (1.3)]. The most complicated
field equations are the ten equations (\ref{216}) which are of Einstein
type relating $W_4$ curvature expressions (on the lhs) to the
energy-momentum tensors of the fields involved in the dynamics (on the
rhs) and an additional coupling term proportional to $\tilde\gamma$.
However, no gravitational constant is identifiable in these scaleless
Weyl-covariant equations.

Moreover, regarding the expression for $\Theta^{(\vphi)}_{\mn}$
we like to point out that the energy-momentum tensor for the $\vphi$
field here automatically appears in the ``new improved'' form including
the terms in the curly brackets (compare Ref.\cite{CCJ}). In our
context these additional contributions expressed in terms of $\Phi^2$,
with Weyl weight $-1$, appear of course in Weyl-covariant form, i.e.\
constructed with Weyl-covariant derivatives. This ``new improved''
addition to the conventional energy momentum tensor, $T_{\mn}^{(\vphi)}$,
for the $\vphi$ field, which was introduced by Callan, Coleman and Jackiw
at first in a $\vphi^4$-theory formulated in flat Minkowski space,
originates here from the variation of the second term in the Lagrangean
(\ref{22}) involving the $W_4$ curvature scalar, i.e.\ from the term
in the variational integral
\begin{equation}
-K\sqrt{-g}{1\over 12}\Phi^2\delta R=-K\sqrt{-g}\ha\Phi^2
\left[(\delta\bar R_{\mn}+\delta P_{\mn})g^{\mn} +(\bar R_{\mn}+P_{\mn})
\delta g^{\mn}\right] . \label{224}
\end{equation}

Let us next check the trace condition following from (\ref{216}).
In order to find the trace of (\ref{216}) we first compute the traces
of the energy momentum tensors (\ref{217}) -- (\ref{219}) for solutions of the
field equations, i.e.\ by using the equations (\ref{213}) -- (\ref{215})
in the derivation. The result is
\begin{eqnarray}
\Theta^{(\vphi)}_\mu{}^\mu &=&-{1\over 6}\Phi^2 R+\tilde\ga\sqrt{\Phi^2}
(\bar\psi\psi)\,,\label{225}\\
T^{(\psi)\mu}_\mu &=& 3\tilde\ga\sqrt{\Phi^2}(\bar\psi\psi)\,,\label{226}\\
T^{(f)\mu}_\mu &=& 0\,.\label{227}
\end{eqnarray}
For the trace of (\ref{216}) one now finds with the help of
(\ref{223}) and (\ref{225}) -- (\ref{227}) that the $\tilde\ga$ term
drops out of this equation and that the rest is identically satisfied
yielding thus no further constraints. This situation may change when we
consider the Weyl-covariant divergence of (\ref{216}) and make use of
the contracted Bianchi identities for a $W_4$ for the geometric quantities.
Then four energy and momentum balance equations are obtained which could
contain additional information worth to be extracted.

In order to compute the Weyl-covariant divergence of (\ref{216})
we need the corresponding divergences of Eqs.(\ref{217}) -- (\ref{219}),
again for solutions of the field equations (\ref{213}) -- (\ref{215}).
Using these as well as the formulae (A33) - (A36) of
Appendix A one finds by a straightforward but lengthly calculation
the following results:
\begin{eqnarray}
D^\mu\Theta^{(\vphi)}_{\mn}~&=&~{1\over 6}
\left[R_{(\nu\r )}-\ha g_{\nu\r}R\right]
D^\r\Phi^2 -{1\over{12}}\Phi^2 D^\r f_{\r\nu}+\tilde\ga (D_\nu
\sqrt{\Phi^2})(\bar\psi\psi)\,,\label{228}\\
D^\mu T^{(\psi)}_{\mn}~&=&-{1\over 4}\left(\bar\psi\left[\ga^\mu\R_{\mn}+\R_{\mn}
\ga^\mu\right]\psi\right)-{1\over 16}\left(\bar\psi\left[\ga_\nu S^{\mu\r}+
S^{\mu\r}\ga_\nu\right]\psi\right)f_{\mu\r} \nonumber\\
& &\qquad + \tilde\ga\sqrt{\Phi^2} D_\nu (\bar\psi\psi )\,,\label{229}\\
D^\mu T^{(f)}_{\mn}~ &=&~6\tilde\a f_{\nu\rho} D^\r R\,,\label{230}
\end{eqnarray}
with $\R_{\mn}$ as defined in Eq. (\ref{253}) below. One can show that the first
two terms in (\ref{229}) each have the form of the coupling of a dual curvature,
${}^*f_{\mn}$, to an axial vector current $(\bar\psi\ga^\nu\ga^5\psi)$
of the spinor field representing
the work done by the dual $f$-curvature field on the axial vector
current of the $\psi$-field given by the expression ${}^*f_{\mn}\,(\bar\psi\ga^\nu
\ga^5\psi)$ times a numerical constant. In fact, in order
to rewrite Eq. (\ref{229}) in a form involving the
hermitean axial vector current $(\bar\psi\ga^\nu\ga^5\psi)$
of the $\psi$ field one has to use Eq. (\ref{254}) below and the formula
\begin{equation}
\left\{\ga^k,S^{ij}\right\}=\ga^k S^{ij}+S^{ij}\ga^k=\ve^{kij}{}_l~
\ga^l\ga^5\/,\label{231}
\end{equation}
where $\ve^{kijl}$ is the Levi-Civita tensor with $\ve_{0123}=+1$, and
\begin{equation}
\ga^5=i\ga^0\ga^1\ga^2\ga^3\quad\mbox{with}\quad 
\ga^5{}^{\dag} =\ga^5\/.
\label{232}
\end{equation}
The dual $f$-curvature tensor is defined by
\begin{equation}
{}^*f_{ik}=\ha\ve_{ikjl} f^{jl}\/\label{233}
\end{equation}
with the corresponding Greek indexed quantities being, as usual,
obtained by conversion with the help of the vierbein fields
yielding ${}^*f_{\nu\r}=\ha\sqrt{-g}~\varepsilon_{\nu\r\mu\si}f^{\mu\si}$
possessing Weyl weight zero. Here the following transformation rule
for the $\varepsilon$-tensor is to be used:
\begin{equation}
\varepsilon_{ijkl}=\l^\mu_i\l^\nu_j\l^\r_k\l^\si_l\sqrt{-g}~
\varepsilon_{\mu\nu\r\si}.
\label{2.33a}
\end{equation}

After these changes in (\ref{229}) one now obtains
with the help of Eqs. (\ref{228}) -- (\ref{230}) and
the twice contracted Bianchi
identities in the form of Eq. (A40) from
Eqs. (\ref{216}) for solutions of the field equations (\ref{213}) -- (\ref{215})
a set of covariant divergence relations which are {\sl identically satisfied}.
Thus no constraints arise in the Weyl-symmetric theory defined by the
Lagrangean (\ref{22}) from the work done by the ${}^*f$-curvature
field on the axial vector current of the spinor field, i.e. the
first two terms in (\ref{229}), in fact,
cancel. Also the contributions of the scalar field disappear from
the divergence relations, and the terms proportional to $\tilde\ga$
resulting from the Yukawa-like coupling between the
$\vphi$- and the $\psi$-field cancel as well in these relations.
Hence the energy and momentum balance in our Weyl-symmetric
massless theory is automatically satisfied for nonvanishing $f$- and
${}^*f$-curvature. This situation changes when the Weyl-symmetry
is broken in Sect. IV.

We, finally, mention that for the Weyl-covariant divergence of the
axial vector current $(\bar\psi\ga_\mu\ga^5\psi)$ of Weyl weight $-1$ one
obtains from the Dirac equation (\ref{214}) and its adjoint the result:
\be
D^\mu (\bar\psi\ga_\mu\ga^5\psi)={1\over\sqrt{-g}}\pa_\mu
(\sqrt{-g} ~\bar\psi\ga^\mu\ga^5\psi)=-2i\tilde\ga\sqrt{\Phi^2}
(\bar\psi\ga^5\psi)\/, \label{249}
\ee
where the first equality is due to the fact that the $\ka_\rho$-contribution
of the connection coefficients (A1) cancels against the
Weyl weight contribution for a quantity of Weyl weight $-1$ [compare
Eq.~({B6})]. Eq.~(\ref{249}) shows that the axial vector
current of the Dirac spinor field has nonvanishing divergence
provided $\tilde\ga\not= 0$.

We conclude this section by considering the square of the Dirac operator
\be
\D =-i\ga^\mu D_\mu\/,\label{250}
\ee
where $\D$ is a matrix-valued operator of Weyl weight $-\ha$. Let us
consider the square of $\D$ applied to a Dirac field $\psi$ of
Weyl weight $w(\psi)=-{3\over 4}$ [compare Schr\"odinger \cite{S} for
the Riemannian case]:
\be
\D\D\psi =-g^{\mn}D_\mu D_\nu\psi - {1\over 4} R\psi + {i\over 4}
f_{\mn}S^{\mn}\psi \/.\label{251}
\ee
To derive Eq.~(\ref{251}) one uses the cyclic identities (A36)
together with the equation
\be
[D_\mu, D_\nu]\psi = i\R_{\mn}\psi - w(\psi) f_{\mn}\psi\,,\label{252}
\ee
where
\be
\R_{\mn}=\pa_\mu\Gamma_\nu -\pa_\nu\Gamma_\mu +i[\Gamma_\mu,\Gamma_\nu]
\label{253}
\ee
is the matrix-valued $W_4$-curvature associated with the spin connection
$\Gamma_\nu$, defined in (\ref{26}), which is related to the curvature
tensor defined in (A19) by
\be
\R_{\mn}=\lambda^k_\mu\lambda^l_\nu \,\ha R_{klij} S^{ij}\,.\label{254}
\ee
To obtain (\ref{251}) one, moreover, needs the relation
\be
D_\r\ga^\mu =\nabla_\r\ga^\mu +\ha\kappa_\r\ga^\mu
+ i[\Gamma_\r,\ga^\mu] =0
\label{255}
\ee
which is analogous to the equation (A9) for the vierbein
fields $\l^\mu_i$ yielding, with (\ref{212}), $D_\r g^{\mn}=0$
[compare (A4)]. The derivation of (\ref{251}) is facilitated
by making use of the Bach tensor [compare Eqs. (A41) -- (A43)]
possessing the same symmetries as the Riemann-Christoffel
tensor $\bar R_{\mu\nu\r\si}$ in a $V_4$.

Using now the Dirac equation (\ref{214}) we compute for the lhs
of (\ref{251})
\be
\D\D\psi =\tilde\ga\left\{-i\ga^\mu\psi D_\mu\sqrt{\Phi^2}+\tilde\ga
\Phi^2\psi\right\}\/.\label{256}
\ee
If we now demand that the squared Dirac operator has a sharp nonzero
eigenvalue originating from the Yukawa-like coupling proportional to
$\tilde\gamma$, and that this eigenvalue is the same
for all components of $\psi$,
we have to conclude that the first term on the rhs of (\ref{256}),
being nondiagonal in spin space, vanishes requiring that $\Phi^2$
is covariant constant, i.e.
\be
D_\mu\Phi^2 =\pa_\mu\Phi^2+\ka_\mu\Phi^2=0\,.
\label{257}
\ee
In this case (\ref{256}) together with (\ref{251}) would read, suppressing
the $f_{\mn}$-term wich is zero in this case (see below),
\be
\D\D\psi =-g^{\mn} D_\mu D_\nu\psi
-{1\over 4} R\psi=\tilde\ga^2\Phi^2\psi
\label{258}
\ee
where, for constant modulus $\Phi$ (see Sect.~IV and V below), $\tilde\ga^2\Phi^2$
would play the role of a quadratic mass term,
$\left({Mc\over\hbar}\right)^2$,
for the $\psi$ field. However, $D_\mu\Phi^2=0$ --~which is a typical
symmetry breaking relation~-- implies according to
(\ref{257})
\be
\ka_\r =-\pa_\r \log\Phi^2\,,
\label{259}
\ee
expressing the fact that the $\ka_\r$-field is ``pure gauge'' and,
hence, the $f$-curvature is vanishing. For this reason we suppressed
above the $f_{\mn}$-term in going from (\ref{251}) to(\ref{258}).
Therefore only in the limiting case of a $W_4$ reducing to a $V_4$ can the
square of the Dirac operator be a diagonal operator with eigenvalue
$\tilde\ga^2\Phi^2$. We shall come back to this property in Sect.~IV
below when we consider the explicit breaking of the Weyl-symmetry.
\section{Massless Weyl-Invariant Description Including Electromagnetism}

In this section we extend the theory presented in Sec.~II and
include the electromagnetic interaction yielding a $D(1)$ {\it and}
$U(1)$ gauge invariant massless theory containing also a dynamics
for the metric in the form of ten Weyl-covariant equations for the
metric tensor $g_{\mn}$ similar to the ones obtained in the
previous section. These field equations (see Eqs. (\ref{310}) below)
are of general relativistic type satisfying the general covariance
principle besides their $D(1)$ and $U(1)$ gauge covariance.
However, as already mentioned above, they contain no gravitational
coupling constant. Such a constant can only be identified after
the Weyl-symmetry has been broken to yield a Riemannian description
in the limit (see Sects. IV and V below).
This is so because Einstein's field equations in general
relativity coupled to material sources are not conformally invariant.

To include electromagnetism in a Weyl-invariant scheme the respective
fiber bundles introduced so far are generalized by adding an
additional $U(1)$ fiber and extending the structural groups of
$P_{\w}$ and $\bar P_{\w}$ to $SO(3,1)\otimes D(1)\otimes U(1)$ and
$Spin (3,1)\otimes D(1)\otimes U(1)$, respectively. We call the
resulting principal bundles $\tilde P_{\w}$ and $\tilde{\bar P}_{\w}$,
respectively, with a pull back of a connection on $\tilde P_{\w}$ being
given by the set of one-forms $(\om_{ik}=-\om_{ki},\ka,A)$ and the
corresponding curvature two-forms ($\Omega_{ik}=-\Omega_{ki},f,F)$,
and analogously for $\tilde{\bar P}_{\w}$. In a natural basis one has,
in addition to the relations of Appendix A:
\be
A=A_\mu dx^\mu\/,\/\qquad F=dA=\ha dx^\mu\wedge dx^\nu F_{\mn}\,,
\label{31}
\ee
with the electromagnetic potentials $A_\mu$ of Weyl weight zero and the
electromagnetic field strengths $F_{\mn}$ [$U(1)$  or $F$-curvature]
of Weyl weight zero, the latter being given by
\be
F_{\mn}=D_\mu A_\nu -D_\nu A_\mu =\pa_\mu A_\nu -\pa_\nu A_\mu\,,
\label{32}
\ee
obeying
\be
D_{\{\lambda}F_{\mn\}} = \pa_{\{\lambda}F_{\mn\}}=0\,.
\label{33}
\ee
In (\ref{32}) and (\ref{33}) the Weyl-covariant derivatives could be
replaced by the ordinary ones due to the symmetry of the connection
coefficients defined in (A1).

The new Lagrangean, called $\tilde\L_{W_4}$, is obtained from
$\L_{W_4}$, defined in (\ref{22}), by the minimal substitution
of the Weyl-covariant derivatives, $D_\nu$, by $D(1)$ {\sl and}
$U(1)$ as well as generally covariant derivatives, $\tilde D_\mu$,
i.e.\
\begin{eqnarray}
D_\mu\vphi\longrightarrow\tilde D_\mu\vphi &=& D_\mu\vphi +
{iq\over\hbar c} A_\mu\vphi\/, \label{34a}\\
D_\mu\vphi^* \longrightarrow \tilde D_\mu\vphi^* &=& D_\mu\vphi^* -
{i q\over\hbar c} A_\mu\vphi^*\/,\label{34b}
\end{eqnarray}
and analogously
\begin{eqnarray}
\orD\psi &\longrightarrow & \overrightarrow{\tilde D}_\mu\psi
= \left(\orD +{ie\over\hbar c} A_\mu \cdot {\bf 1}\right)\psi\,,\label{35a}\\
\bar\psi\olD &\longrightarrow & \bar\psi\overleftarrow{\tilde D}_\mu
=\bar\psi \left(\olD -{ie\over\hbar c} A_\mu \cdot {\bf 1}\right)\,.
\label{35b}
\end{eqnarray}
Here we have denoted the charge of the field $\vphi$ by $q$ and
that of $\vphi^*$ by $-q$, while the charge of the fermion field $\psi$
is denoted by $e$ and that of its adjoint by $-e$. In addition to the
substitutions (\ref{34a}) -- (\ref{35b}) we have to add, in the familiar way,
a contribution for the free electromagnetic fields $F_{\mn}$
in the Lagrangean $\tilde \L_{W_4}$.

We remark in passing that in the minimal electromagnetic interaction,
by introducing electromagnetism
through the substitutions (\ref{34a}) -- (\ref{35b}) into a supposedly
known system of fields in quantum field theory, one usually regards
the potentials $A_\mu$ and the corresponding fields $F_{\mn}$ as
{\it external fields}. In the present case, however, there are no
external fields the sources of which are not included in the system.
We want to consider a {\sl closed system} in which the electromagnetic
fields are generated by currents of the $\vphi$ and $\psi$ fields,
respectively.
The potentials $A_\mu$ thus describe besides the electromagnetic action
of the field $\vphi$ onto $\psi$ and vice versa also the back reaction
onto their own motion. We shall not investigate these back reactions in
detail here. For a thorough discussion of the self-field contributions
in the relativistic theory of classical charged point particles we refer
to the book by Rohrlich \cite{R}.

Having said that the introduction of electromagnetic interactions should
lead to a closed system of fields described, moreover, in the proposed
Weyl-covariant manner, we hasten to add that it is certainly not
without problems to introduce electromagnetic fields into a massless
theory where the $A_\mu$ fields are coupled to source currents of
{\it massless} $\vphi$ and $\psi$ fields. It is very likely that a
nonsingular description is only possible for $q=e=0$ in a massless
theory. Our excuse for nevertheless introducing electromagnetism
in the standard manner into our Weyl-invariant theory developed in
Sect.~II is that we, ultimately, intend to break the Weyl symmetry
explicitly in order to generate masses for the $\vphi$ and $\psi$
fields. In this scenario the extension of a broken Weyl theory
without electromagnetism to a broken Weyl theory including
electromagnetism is correctly given by the standard
arguments exhibited by the substitutions (\ref{34a}) -- (\ref{35b}).
After these remarks we base our further discussion in this section
on the following Weyl- and $U(1)$-invariant Lagrangean density:
\begin{eqnarray}
\tilde\L_{W_4}=&&K\sqrt{-g}~\Bigl\{\ha g^{\mn}
\tilde D_\mu\vphi^*\tilde D_\nu\vphi
-{1\over 12} R\,\vphi^*\vphi -\beta(\vphi^*\vphi)^2
+\tilde\a R^2 +{i\over 2}
(\bar\psi\ga^\mu\overrightarrow{\tilde D}_\mu
\psi - \bar\psi\overleftarrow{\tilde D}_\mu \ga^\mu\psi)\nonumber\\
&&\qquad\qquad +\tilde\ga\sqrt{\vphi^*\vphi}\,(\bar\psi\psi)-\tilde\d
{1\over 4} f_{\mn}f^{\mn}
-{1\over K}{1\over 4} F_{\mn}F^{\mn}\Bigr\}\,.\label{36}
\end{eqnarray}

In the usually adopted units in which ${e\over\hbar c}A_\mu$ has the
dimension of an inverse length the term ${1\over 4} F_{\mn} F^{\mn}$
has the dimension of an energy/volume. In order to convert this into
a quantity of dimension $L^{-2}$, like all the other terms in the
curly brackets of (\ref{36}), one has to multiply the last term by a
factor $K^{-1}$ (compare in this context the discussion in Sect.~II
above).

The Weyl- and $U(1)$-covariant field equations following from a variational
principle based on the Lagrangean
(\ref{36}) are now given by:
\begin{eqnarray}
\d\vphi^*:~&& g^{\mn}\tilde D_\mu\tilde D_\nu\vphi +{1\over 6} R\vphi+
4\beta (\vphi^*\vphi)\vphi -\tilde\ga {\vphi\over\sqrt{\Phi^2}}(\bar\psi
\psi)=0\,,\label{37}\\
\d\psi^{\dagger }:~&&-i\ga^\mu\tilde D_\mu\psi -
\tilde\ga\sqrt{\Phi^2}\psi=0\,,\label{38}\\
\d\ka_\r:~&&\tilde\d D_\mu f^{\mu\r}=-6\tilde\a D^\rho R\,, \label{39}\\
\d g^{\mn}:~&&{1\over 6}\Phi^2\left[R_{(\mn)}-\ha g_{\mn} R\right]-4\tilde\a R
\left[R_{(\mn)}-{1\over 4}g_{\mn} R\right]-4
\tilde\a\Bigl\{D_{(\mu}D_{\nu)}R-
g_{\mn}D^\r D_\r R\Bigr\}=\nonumber\\
&&\qquad\qquad\qquad =\tilde\Theta_{\mn}^{(\vphi)}+\tilde T_{\mn}^{(\psi)}+
T_{\mn}^{(f)}+T^{(F)}_{\mn}-g_{\mn}\tilde\ga\sqrt{\Phi^2}
(\bar\psi\psi)\,,\label{310}\\
\d A_\r:~&& D_\mu F^{\mu\r}={K\over\hbar c}
\left[q j^{(\vphi)\r}+e j^{(\psi)\r}\right]\,.
\label{311}
\end{eqnarray}
Here the $D(1)$ and $U(1)$ gauge covariant hermitean currents are
defined by (pulling the index $\rho$ down with $g_{\r\mu}$):
\begin{eqnarray}
j^{(\vphi)}_\mu &=& {i\over 2} [\vphi^*\cdot\tilde D_\mu \vphi - \tilde D_\mu
\vphi^*\cdot\vphi]\,,\label{312}\\
j^{(\psi)}_\mu &=& (\bar\psi\ga_\mu\psi)\,.\label{313}
\end{eqnarray}
Both currents possess Weyl weight $-1$. Remember that the Weyl weight of
$\psi$ was chosen to be $w(j^{(\psi)}_\mu)=-1$ so that the
Dirac vector current (\ref{313}) can act as a source current in
Maxwell's equations (\ref{311}), where again we can replace the lhs
by [compare (\ref{222})]
\be
D_\mu F^{\mu\r}=\bar\nabla_\mu F^{\mu\r}={1\over\sqrt{-g}}\pa_\mu
(\sqrt{-g}~F^{\mu\r})\label{314}
\ee
being thus a quantity of Weyl weight $-1$ independent of $\kappa_\r$.
The $\ka_\r$-contributions also disappear from the rhs of (\ref{312})
with only the electromagnetic contributions remaining as in the usual
Klein-Gordon theory with electromagnetic interaction. Eq.~(\ref{39})
and the definition of $T^{(f)}_{\mn}$ are unchanged
[see Eqs.~(\ref{215}) and (\ref{219})], while
$\tilde\Theta_{\mn}^{(\vphi)},\tilde T_{\mn}^{(\psi)}$
and $T^{(F)}_{\mn}$ are given by the following $D(1)$ {\sl and} $U(1)$
gauge covariant expressions symmetric in $\mu$ and $\nu$:
\begin{eqnarray}
\tilde\Theta^{(\vphi)}_{\mn}=~&&\ha (\tilde D_\mu \vphi^*\tilde D_\nu\vphi +
\tilde D_\nu\vphi^*\tilde D_\mu\vphi)-{1\over 6}\Bigl\{D_{(\mu} D_{\nu)}
\Phi^2-g_{\mn} D^\rho D_\rho\Phi^2\Bigr\}\nonumber\\
&&\qquad\qquad\qquad  -g_{\mn}\left[\ha g^{\rho\lambda}
\tilde D_\r\vphi^*\tilde D_\l\vphi -\beta
(\vphi^*\vphi)^2\right]\,, \label{315}\\
\tilde T_{\mn}^{(\psi)} =~&&{i\over 2}\left\{\bar\psi\ga_{(\mu}
\overrightarrow{\tilde D}_{\nu)}\psi -\bar\psi
\overleftarrow{\tilde D}_{(\mu}\ga_{\nu)}\psi\right\}-
g_{\mn}{i\over 2}\left\{\bar\psi\ga^\rho\overrightarrow{\tilde D}_\rho\psi -
\bar\psi\overleftarrow{\tilde D}_\r\ga^\r\psi\right\}\,,\label{316}\\
T_{\mn}^{(F)} =~&& -{1\over K} \left[F_{\mu\si} F_\nu {}^\si -{1\over 4} 
g_{\mn} F^{\r\lambda} F_{\r\l}\right]\,.\label{317}
\end{eqnarray}

The total electromagnetic current appearing as source current on the
rhs of (\ref{311}) is conserved as a consequence of (\ref{311}) and the
relation $D_\r D_\mu F^{\mu\r}=0$. On the other hand, using (\ref{37}) and its
complex conjugate, one concludes from (\ref{312}) that the 
$\vphi$-current
alone is conserved. Similarly, one concludes from (\ref{38}) and its
adjoint that the vector current (\ref{313}) is conserved, i.e.\ we
have the two separate charge conservation equations
\begin{eqnarray}
D^\mu j^{(\vphi)}_\mu &=&{1\over\sqrt{-g}}\pa_\mu (\sqrt{-g}~g^{\mu\r}
j^{(\vphi)}_\r)=0\,,\label{318}\\
D^\mu j^{(\psi)}_\mu &=& {1\over\sqrt{-g}}\pa_\mu (\sqrt{-g}~g^{\mu\r}
j^{(\psi)}_\r) =0\,,\label{319}
\end{eqnarray}
where the $\ka_\r$-contributions in the connection coefficients (A1)
and the Weyl weight term cancel for a vector of Weyl weight $-1$ as
mentioned before.\\
The axial vector current of the Dirac field again satisfies
the divergence relation (\ref{249}) for the theory including
electromagnetism. Moreover, since Eq. (\ref{39}) is the same as
(\ref{215}), also the relation (\ref{223}) remains valid in the
theory based on the Lagrangean $\tilde\L_{W_4}$. Clearly, there is no
coupling between the two gauge fields $\ka_\r$ and $A_\r$ for
a gauge theory which is of direct product type.

The discussion of the trace of Eqs. (\ref{310}) leads to the same
result as in Sect. II since the additionally appearing trace of the
electromagnetic energy-momentum tensor (\ref{317}) is zero:
\be
T_\mu^{(F)\mu}=0\,.\label{320}
\ee
We, finally, discuss the divergence relations 
following from Eqs. (\ref{310})
including electromagnetic interactions. Using the following formula
for the commutator of the covariant derivatives $\tilde D_\mu$, analogous
to Eqs. (A34) and (A35) but including electromagnetism:
\be
[\tilde D_\mu,\tilde D_\nu]=[D_\mu,D_\nu]+{ie\over\hbar c} F_{\mn}\,,
\label{321}
\ee
one now finds instead of Eqs. (\ref{228}) and (\ref{229}) -- the
latter after computing the again vanishing axial vector coupling
resulting from the first two terms on the rhs -- for
the theory based on $\tilde\L_{W_4}$, and for the solutions of
the field equations (\ref{37}) -- (\ref{311}), the following relations:
\begin{eqnarray}
D^\mu\tilde\Theta^{(\vphi)}_{\mn}&=&{1\over 6}
\left[R_{(\nu\r)}-\ha g_{\nu\r} R\right]
D^\r\Phi^2-{1\over 12}\Phi^2 D^\r f_{\r\nu}
+\tilde\gamma (D_\nu\sqrt{\Phi^2})(\bar\psi\psi)+{q\over\hbar c}
F_{\nu\r}~j^{(\vphi)\r}\,,\label{322}\\
D^\mu\tilde T_{\mn}^{(\psi)} &=&
\tilde\ga\sqrt{\Phi^2}D_\nu (\bar\psi\psi)+
{e\over\hbar c}F_{\nu\r}~j^{(\psi)\r}\,,\label{323}
\end{eqnarray}
with (\ref{230}) remaining unchanged, i.e.
\be
D^\mu T_{\mn}^{(f)}= 6\tilde\a f_{\nu\r} D^\r R\,,\label{324}
\ee
and
\be
D^\mu T_{\mn}^{(F)} = -F_\nu{}^\r \left[{q\over\hbar c} j^{(\vphi)}_\r
+{e\over\hbar c} j^{(\psi)}_\r\right]\,.\label{325}
\ee
Here and in Eqs. (\ref{322}) and (\ref{323}) the currents $j^{(\vphi)}_\r$
and $j^{(\psi)}_\r$ are defined by Eqs. (\ref{312}) and (\ref{313}),
respectively, 
while $\tilde\Theta^{(\vphi)}_{\mn}, \tilde T^{(\psi)}_{\mn}$
and $T^{(F)}_{\mn}$ are defined in Eqs.(\ref{315}), (\ref{316}) and
(\ref{317}), respectively. With the help of Eqs. (\ref{322}) -- (\ref{325})
the divergence relations following from (\ref{310}), including
electromagnetic effects, are now again identically satisfied showing that
also the electromagnetic contributions appearing on the rhs
of Eqs. (\ref{322}), (\ref{323}), and (\ref{325}) cancel in these energy
and momentum balance relations implying
that one is, indeed, considering an electromagnetically closed system
of fields.

Computing finally again the square of the Dirac operator $\tilde{\D} =
-i\ga^\mu\tilde D_\mu$ one obtains using
(\ref{321}) together with (\ref{252}) along the same lines as in
Sect. II the $U(1)$ and Weyl-covariant result for $w(\psi)=-{3\over 4}$:
\be
\tilde{\D}\tilde{\D}\psi =-g^{\mn}\tilde D_\mu\tilde D_\nu\psi -
{1\over 4} R\psi + {i\over 4} f_{\mn} S^{\mn}\psi
- {e\over\hbar c} F_{\mn} S^{\mn}\psi\,.
\label{327}
\ee
\section{Weyl-Symmetry Breaking}

Having formulated the theory of a massless scalar field and a massless
fermion field in a Weyl space $W_4$ in the presence of electromagnetic
fields, we now break the Weyl-symmetry by a term in the Lagrangean
constructed with the help of the curvature scalar $R$ of a $W_4$ and a
mass term for the $\vphi$ field regarding the modulus of the
scalar field as a Higgs-type field. Due to the
Yukawa-like coupling proportional to $\tilde\ga$ of the $\vphi$- and the
$\psi$-fields there appeared already in the Weyl-symmetric theory
treated in Sects. II and III a mass-like term for the fermion field
in the field equations for $\psi$ [see Eqs. (2.14) and (3.10)]
which, in the broken theory, when
the $W_4$ reduces to a $V_4$, will yield a mass term proportional
to $\tilde \ga$ for the $\psi$ field. The total Lagrangean density 
on which we shall base the discussion
in this section is thus given by
\be
\L =\tilde{\L}_{W_4}+\L_B\;,
\label{41}
\ee
where $\tilde{\L}_{W_4}$ is the Weyl-invariant Lagrangean defined
in (\ref{36}) and $\L_B$ is the $U(1)$ gauge invariant but
Weyl-symmetry breaking term of Weyl weight +1 defined by
\be
\L_B=-{a\over 2} K\sqrt{-g}\left\{{1\over 6} R+\left[{mc}\over{\hbar}
\right]^2\vphi^*\vphi\right\}\;.
\label{42}
\ee
The constant $a$ has the length dimension $[L^0]$. The symmetry
breaking Lagrangean (\ref{42}) introduces a length scale into the
theory given by the Compton wave length associated with the mass of
the $\vphi$ field, and this scale breaking is related to the geometry of the
underlying Weyl space being determined by the scalar curvature $R$ of the
$W_4$ defined in (A31). The Weyl geometry thus provides the
geometric framework for the breaking mechanism introducing nonzero
masses for the scalar field which itself is not to be regarded as a
true, bona fide, matter field. We shall see that the modulus
$\Phi$ of $\vphi$ behaves like a geometric quantity in this
formalism with $\Phi^2$ acting as a potential for the Weyl vector
fields $\ka_\r$. The role of the geometry in the mass giving procedure as
described here is that of an embedding stratum or medium
which takes part in the
phenomenon. We thus refer to this form of mass generation for short as
the ``Archimedes principle'', since the ambient geometry is taking
an active part in it.

Let us now write down the field equations following from a variational
principle formulated with the Lagrangean $\L$. One finds, using the
same notation as before:
\begin{eqnarray}
\d\vphi^*:~ && g^{\mn}\tilde D_\mu\tilde D_\nu \vphi+{1\over 6} R\vphi +
4\beta (\vphi^*\vphi)\vphi -\tilde\gamma{\vphi\over\sqrt{\Phi^2}}
(\bar\psi\psi)+ a\left[{mc\over\hbar}\right]^2\vphi =0\;,\label{43}\\
\d\psi^\dagger :~ && -i\gamma^\mu \tilde D_\mu\psi 
-\tilde\ga\sqrt{\Phi^2}\psi =0\;,
\label{44}\\
\d\kappa_\r :~ &&\tilde\d D_\mu f^{\mu\r}=-6\tilde\a D^\rho R+{a\over 4}
\ka^\r\;,\label{45}\\
\d g^{\mu\nu}:~&&{1\over 6}(\Phi^2+a)\left[R_{(\mn)}-{1\over 2}
g_{\mn} R\right] -
    4\tilde\a R\left[R_{(\mn)}-{1\over 4} g_{\mn} R\right]
-4\tilde\a \left\{D_{(\mu}D_{\nu)} R-g_{\mn} D^\rho D_\rho R\right\}=
\nonumber \\
&&\qquad =\tilde\Theta^{(\vphi)}_{\mn}+\tilde T_{\mn}^{(\psi)} +
T_{\mn}^{(f)} +T_{\mn}^{(F)}-g_{\mn}\tilde\ga\sqrt{\Phi^2}(\bar\psi\psi)
+g_{\mn}{a\over 2}\left[{mc\over\hbar}\right]^2\Phi^2\;,\label{46} \\
\d A_\r:~ && D_\mu F^{\mu\rho}={K\over \hbar c}~\left[qj^{(\vphi)\rho}+
ej^{(\psi)\rho}\right]\;,\label{47}\\
\d  a:~ &&{1\over 6} R+\left[{mc\over\hbar}\right]^2\Phi^2 =0\;.\label{48}
\end{eqnarray}

Computing now first the traces of $\tilde\Theta^{(\vphi)}_{\mn}$ and
$\tilde T^{(\psi)}_{\mn}$ using the field equations (\ref{43}) and
(\ref{44}), respectively, one finds
\be
\tilde\Theta^{(\vphi)\mu}_{\mu} = -{1\over 6}\Phi^2 R +\tilde\ga
\sqrt{\Phi^2}(\bar\psi\psi) - a\left[{mc\over\hbar}\right]^2\Phi^2\;,
\label{49}
\ee
and
\be
\tilde T_\mu^{(\psi)\mu}= 3\tilde\ga \sqrt{\Phi^2}~(\bar\psi\psi )\;.
\label{410}
\ee
With these results one obtains from (\ref{46}) by contracting
this equation with $g^{\mn}$ and using the field equations as well as
the constraint (\ref{48}) and Eqs.~(\ref{227}) and (\ref{320}) that
\be
\tilde \a D^\rho D_\rho R= 0\;,\label{411}
\ee
which is an equation we had obtained before in Sects. II and III from
the $\ka_\rho$-equations [compare Eq.~(\ref{223})].
Now we conclude from (\ref{45}) by taking the Weyl-covariant divergence
and using (\ref{411}) that, for $a\not= 0$, the Weyl vector fields must
satisfy the Lorentz-type condition
\be
D^\rho\ka_\rho =0\;.\label{412}
\ee
This equation may, with the help of (B6), be turned into the following
$V_4$ covariant form:
\be
\bar\nabla^\rho\ka_\rho ={1\over\sqrt{-g}}\pa_\r(\sqrt{-g}~\ka^\r)=
\ka_\rho\ka^\rho\;.\label{413}
\ee
Eq. (\ref{413}) implies that the $W_4$ curvature scalar may now be
expressed as
\be
R=\bar R-{3\over 2}\ka_\rho\ka^\rho\;,\label{414}
\ee
and $R_{(\mn)}$ is given by
\be
R_{(\mn)} =\bar R_{\mn}+P_{(\mn)}=\bar R_{\mn}-{1\over 2}
(\bar\nabla_\mu\ka_\nu +\bar\nabla_\nu\ka_\mu)-\ha \ka_\mu\ka_\nu\;.
\label{415}
\ee

Let us now turn to the divergence relations following from (\ref{46})
for the solutions of the field equations. We know from our previous
discussions in Sect.~II and III that these energy and momentum balance
relations, obtained after using the contracted Bianchi identities, were
satisfied identically in the unbroken theory.
Collecting now the $a$-dependent terms in the covariant divergence
of Eqs. (\ref{46}), and remembering that the $\vphi$-equation (\ref{43})
now contains a mass term, yields for the divergence relations following
from (\ref{46}) for the solutions of the field equations 
(\ref{43}) -- (\ref{45}) the result:
\begin{equation}
{a\over 3} D^\mu f_{\mn} - a f_{\nu\mu} \ka^\mu = 0
\;.\label{416}
\end{equation}
These relations are trivially satisfied for $D_\nu\Phi^2 = 0$ implying
$f_{\mn}=0$; i.e. for $\ka_\nu = -\pa_\nu\log\Phi^2$ being ``pure gauge''
(compare the discussion at the end of Sect. II concerning the
square of the Dirac operator). Hence Eqs. (\ref{416})
are satisfied in the limit of the Weyl space $W_4$ being equivalent to
a Riemannian space $V_4$. Using now the field equations (\ref{45}) for
vanishing $f$-curvature, together with Eq. (\ref{48}) and $D_\mu\Phi^2=0$,
one easily concludes that
\be
\ka_\r = 0\;, \mbox{~~i.e. that~~} \Phi^2 = const \;,
\label{417}
\ee
implying that the Weyl space $W_4$, in fact, reduces completely
to a Riemannian space $V_4$.

Considering, finally, the square of the Dirac operator for the case 
(\ref{417}) [compare Eqs. (2.44) and (3.29)] one sees that it is
a diagonal operator with constant eigenvalue $\tilde\ga^2\Phi^2$.

Summarizing we can say that with
$a\neq 0$ and the Weyl-symmetry breaking relation $D_\mu\Phi^2=0$,
with $\Phi^2 = const$, we
have to view Eqs. (\ref{43}) -- (\ref{48}) as a set of covariant field
equations formulated in a $V_4$ with a definite metric (see Sect. V below).
Both the $\vphi$-field and the $\psi$-field have now acquired a mass
in this quasi classical (i.e. single-particle) theory with the mass
of the spinor field being determined by the modulus $\Phi $ of the
scalar field. This result is reminiscent of the situation occuring in the
Higgs phenomenon of a nonabelian gauge theory (for example the
standard model) where the masses of the (nonabelian) gauge and
fermion fields are determined by the vacuum expectation value of the
scalar field possessing a nonlinear selfcoupling of the same type
as discussed here, i.e. by the {\sl classical part}
contained in the quantized scalar field of the model. Since
the gauge sector --~after Weyl symmetry breaking~-- contains in the
present case only the massless $A_\mu$-fields [the $U(1)$ gauge
fields] and classical gravitation [the Lorentz gauge fields in a
vierbein formulation] the Higgs-type mass giving phenomenon only
manifests itself, besides for the $\vphi$-field, in the mass term
for the $\psi$-field, obtained for constant $\Phi^2$, implying in
turn the reduction of the Weyl space to a Riemannian space.

\section{Determination of the Free Parameters}

Having broken in the last section the original massless Weyl and
$U(1)$ symmetric formulation of the theory to a $U(1)$ gauge
theory of interacting massive, single-particle, scalar and spinor
fields formulated in a pseudo-Riemannian space $V_4$, i.e. taking
gravitational effect into account, we now have to determine the
free parameters of the theory and relate the description to the
standard one and, in particular, see how Einstein's equations for
the metric appear on the scene.

Considering Eqs. (\ref{43}) - (\ref{48}), as discussed for the case
$D_\r\Phi^2=0$ with $\Phi^2=const > 0$, and hence $\ka_\r=0$,
yields from (\ref{48})
\be
R=\bar R= -6\left[{mc}\over {\hbar}\right]^2\Phi^2\,.
\label{51}
\ee
Thus the underlying space $V_4$ is of constant negative
curvature being isomorphic to the noncompact coset space \cite{H}
$G/H=O(1,4)/O(1,3)$. Furthermore, with $\ka_\r=0$ Eq. (\ref{45})
reads $\pa_\r\bar R=0$ in accord with (\ref{51}).

The field equations for $\vphi=\Phi e^{{i\over\hbar}S}$ (with
constant modulus) and $\psi$ now read:
\begin{eqnarray}
g^{\mn}\tn_\mu\tn_\nu\vphi +\left\{\left[{mc}\over{\hbar}\right]^2
(a-\Phi^2)+4\beta\Phi^2-\tilde\ga {{1}\over{\sqrt{\Phi^2}}}(\bar\psi\psi)
\right\}\vphi=0\,,
\label{52}\\
-i\ga^\mu\tn_\mu\psi - \tilde\ga\sqrt{\Phi^2}\psi =0\,,
\label{53}
\end{eqnarray}
where we denote by $\tn_\mu \vphi$ and $\tn_\mu \psi$ the metric
and $U(1)$ covariant derivatives (3.4) and (3.6) for
$\ka_\r =0$, respectively.

Regarding the field equations (\ref{46}) in the present case,
we have to remember that the term proportional to $\tilde\a$ was
introduced in the Lagrangeans (\ref{22}) and (3.8) above in
order to yield a nontrivial dynamics for the Weyl vector field $\ka_\r$.
Since the Weyl vector field has disappeared after the symmetry breaking
$W_4\longrightarrow V_4$, considering, moreover, constant $\Phi $, we
may now take $\tilde\a=0$ and rewrite (\ref{46}) as
\be
\bar R_{\mn} - \ha g_{\mn}\bar R = 
{{1}\over{{1\over 6}\left[\Phi^2+a\right]K}}
\left\{\tilde\Theta^{(\vphi)}_{\mn}{}'+\tilde T^{(\psi)}_{\mn}{}'
+ T^{(F)}_{\mn}{}'\right\}\,.
\label{54}
\ee
The primed tensors on the rhs of (\ref{54}) are the
energy-momentum tensors for the {\sl massiv} $\vphi$ and
$\psi$ fields given -- together with $T^{(F)}_{\mn}{}'$ -- by
\begin{eqnarray}
\tilde\Theta^{(\vphi)}_{\mn}{}' &=& K\left[\tilde\Theta^{(\vphi)}_{\mn}
\vert {}_{\ka_\r =0} + g_{\mn}{a\over 2}\left[{mc}\over {\hbar}\right]^2
\Phi^2\right]\,,\label{55}\\
\tilde T^{(\psi)}_{\mn}{}' &=& K\left[\tilde T^{(\psi)}_{\mn}
\vert {}_{\ka_\r =0} + g_{\mn}{{mc}\over {\hbar}}\sqrt{\Phi^2}(\bar\psi\psi)
\right]\,,\label{56}\\
T^{(F)}_{\mn}{}' &=& K~T^{(F)}_{\mn}\,,\label{57}
\end{eqnarray}
where the squared bare gravitational mass of the $\vphi$-field
is $a^2m^2$, and in the mass term of the $\psi$-field we have taken
\be
-\tilde\ga\sqrt{\Phi^2} = {{mc}\over {\hbar}}\sqrt{\Phi^2}
= {{Mc}\over {\hbar}}\,.
\label{58}
\ee
In accordance with (\ref{53}) we have here adopted for the value
of the constant $\tilde\ga$ of dimension $[L^{-1}]$ characterizing the
Yukawa coupling the value $\tilde\ga =-mc/\hbar$
yielding a fermion mass of the correct sign and of the size
$M=m\sqrt{\Phi^2}$ in Eq. (\ref{53}).

In order to fix a conventional mass term in the dynamics
of the $\vphi$-field we choose $a=1$ thereby introducing the
length scale
\be
l_{\vphi} = {{\hbar}\over {mc}}
\label{58a}
\ee
into the theory after symmetry breaking. We shall use this length
scale as the intrinsic unit for measuring the constants
appearing in the Lagrangean (\ref{41}) and in the field equations
derived from it. Thus Eq. (\ref{58}) implies that $\tilde\ga =
-1\cdot l^{-1}_{\vphi}$. Similarly we measure the nonlinear
coupling constant $\beta$ in units of $l^{-2}_{\vphi}$ and write
$\beta =\beta ' l^{-2}_{\vphi}$. The conversion between mass and
length in Eq. (\ref{58a}) is done by assuming $\hbar$ and $c$ to be
given constants of nature characterizing quantum mechanics and
special relativity, respectively.

Now the effective mass squared in units of $l^{-2}_{\vphi}$
of the interacting $\vphi$-field
is  represented by the first three (constant) terms in the curly
brackets in Eq. (\ref{52}) given, with $a=1$ and (\ref{58a}), by
\be
V(\Phi^2) = \left[{mc}\over {\hbar}\right]^2 (a - \Phi^2) +
4\beta\Phi^2 = l^{-2}_{\vphi}\left[1-\Phi^2+4\beta '\Phi^2\right]\,.
\label{59}
\ee
This expression is different from the mass contribution to the
erergy-momentum tensor of the $\vphi$-field which is given by the
second term on the rhs of (\ref{55}) for $a=1$ accounting for the change
in the energy-momentum tensor $\tilde\Theta^{(\vphi)}_{\mn}$ due to
the breaking of the Weyl symmetry induced by $\L_B$. Of course, there
also contributes a term proportional to the nonlinear coupling constant
$\beta$ (assumed to be positive) given by $g_{\mn}\beta (\Phi^2)^2$
[compare the last term in (3.17)] which is contained in
$\tilde\Theta^{(\vphi)}_{\mn}\vert{}_{\ka_\r=0}$ on the rhs of
(\ref{55}) surviving in the limit $D_\r\Phi^2 = \pa_\r\Phi^2 = 0$,
and appearing thus as a source term on the rhs of Eq. (\ref{54}).
Finally, the constant $K$ in Eqs. (\ref{54}) and (\ref{55}) -- (\ref{57})
assures that all the source terms in the field equations (\ref{54}),
i.e. the primed energy-momentum tensors in the curly brackets, 
possess the dimension energy/volume [compare the discussion in Sect. II].
For electromagnetism this implies only a trivial change as given
in (\ref{57}) [see the discussion after Eqs. (3.8) and (3.19) above].

It is easy to show using Eqs. (3.24), (3.25) and (3.27) as well as the
symmetry breaking relation $D_\r\Phi^2=0$, with $\Phi^2=const$, that
the rhs of (\ref{54}) obeys
\be
\bar\nabla\left[\tilde\Theta^{(\vphi)}_{\mn}{}' +
\tilde T^{(\psi)}_{\mn}{}' + \tilde T^{(F)}_{\mn}{}'\right] = 0\,
\label{510}
\ee
expressing the covariant energy-momentum conservation in a $V_4$
with sources provided by the massive, charged, interacting fields
$\vphi$ and $\psi$ together with electromagnetism.

The equations (\ref{54}) are identical with Einstein's field
equations for the metric in general relativity coupled to an
interacting system of massive scalar, massive Dirac-spinor and
massless electromagnetic fields
provided we can identify the factor in front of the curly brackets
with Einstein's gravitational constant 
$\kappa_{\hbox {\tiny E}}=8\pi N/{c^4}=
2.076 \cdot 10^{-48}g^{-1}cm^{-1}sec^2$, where $N$ is Newton's
constant, i.e. provided
\be
\kappa_{\hbox {\tiny E}} =
{{1}\over {{1\over 6}\left[\Phi^2 + 1\right]K}} > 0\,.
\label{511}
\ee

It appears that the over all size of the gravitational coupling
constant (\ref{511}) is determined by the constant $K$ while
$\Phi^2$ is a constant of order unity. To see what the
contribution of $\Phi^2$ relative to the choice $a=1$ may be let us
assume for the moment that there is no fermion field present
and that for a coordinate independent solution of (\ref{52})
to exist one would have to require the vanishing of $V(\Phi^2)$
as defined in (\ref{59}). This would yield
\be
a = 1 = \left(1 - 4\beta '\right) \Phi^2\,.
\label{512}
\ee
relating thus $\Phi^2$ to the value of the constant $\beta$
measured in units of $l^{-2}_{\vphi}$.
For the constant $\ka_{\hbox {\tiny E}}$ this would imply that
in the case of a universal coordinate independent solution
of the $\vphi$-field equation one would obtain for
Einstein's gravitational constant the result
\be
\ka_{\hbox {\tiny E}} = {{1}\over {{1\over 6}\left[{{2 - 4\beta '}\over
{1 - 4\beta '}}\right]K}}~.
\label{513}
\ee
This expression remains positive, for positive K,
for any nonlinear universal
coupling constant $\beta '>0$. We thus obtain in this
theory a universal gravitational coupling constant of the type
(\ref{511}) or (\ref{513}) as in a Brans-Dicke-like scalar-tensor
theory. In the following we shall choose (\ref{511}) as the 
identification of the gravitational constant in the present theory
which is an expression independent of the value for the effective mass
of the $\vphi$-field obtained from (\ref{52}).

Moreover, we have to remember that Eqs. (\ref{54}), with the
specifield source terms indicated, should yield a $V_4$
of constant curvature according to Eqs. (\ref{51}) and (\ref{45})
[the latter for $\ka_\r =0$]. Computing the trace of (\ref{54})
one convinces oneself with the help of (\ref{49}) and (\ref{410}),
using also (\ref{51}) again, that this is consistent with the
field equations for the metric.

So far we have considered a closed system of interacting massive
$\vphi$, $\psi$ and (massless) electromagnetic fields in interaction
with the metric of the underlying space $V_4$ which turned out
to be of constant curvature. Having obtained Einstein's field
equations (\ref{54}) with the gravitational coupling constant
defined by (\ref{511}), we could now extend the
description and formulate the theory in the presence of a classical
background gravitational field -- say of cosmological origin --
by adding another (classical) source term in the curly brackets
on the rhs of (\ref{54}) and drop the requirement that the
underlying space be of constant curvature. 

Another extension of the theory investigated in this
paper would be the extension to a multi-particle theory, in
particular, by considering second quantized spinor fields in this broken
Weyl theory. We shall make some remarks in this direction 
in Sect. VI below.
However, it appears that the complex scalar field $\vphi$
plays a special role in this formalism with its modulus
remaining a classical field. It is, therefore, doubtful whether
a fully quantized scalar field would lead in this context to an
improved formulation of this theory.

We close this section by remarking that the essential parameter
appearing after the breaking of the Weyl symmetry by $\L_B$
defined in (\ref{42}) is, with $a=1$, the mass $m$ of the scalar
field, or rather the length $l_{\vphi}$ defined in (\ref{58a}).
This length scale was used above also as the intrinsic unit in
which the dimensional constants $\tilde\ga$ and $\beta$
appearing already in the Weyl-symmetric theory are ultimately
to be measured. In order to fix the length scale $l_{\vphi}$
to a particular value we could now arbitrarily identify
$m$ with the mass of the $\pi^0$-meson marking the lower edge 
of the hadronic mass spectrum. The value of the constant
$\Phi^2$, finally, determines, according to Eq. (\ref{58}),
the deviation of the mass $M$ of the Dirac field from the
mass of the scalar field.

\section{Discussion}

We have used in this paper a Weyl geometry to formulate the
dynamics of a massless scalar and a massless Dirac spinor field
in the presence of electromagnetic and metric fields with all
the field quantities --~possessing nontrivial Weyl weights~-- being
determined up to conformal rescalings, (1.2) and (A2), coupled to the
corresponding transformations (1.3) of the Weyl vector fields
$\ka_\r$. Then we explicitly broke this Weyl-invariant theory
defined by the Lagrangean $\tilde\L_{W_4}$ by adding a term $\L_B$
constructed with the help of the curvature scalar of the
$W_4$ and a mass term for the $\vphi$-field in order to study the
appearance of a mass term for the fermion field and a
corresponding length scale characterizing the dynamics after
breaking the symmetry of the embedding space from a Weyl space
$W_4$ to a pseudo-Riemannian space $V_4$.

The appearance of Einstein's equations with sources given by the
energy-momentum tensors of the now massive scalar and spinor
fields, together with electromagnetic fields, was investigated
and a gravitational coupling constant was identified by the expression
(5.12). The symmetry reduction from a Weyl space to a Riemannian space
was governed by the equations $D_\r\Phi^2=0$, which was deduced from the
requirement of the square of the Dirac operator to be a diagonal operator
in spin space in the massive case, and from the demand to satisfy the
divergence relations following from the field equations for the metric
together with the contracted Bianchi identities for a $W_4$.
This implies the relations $\ka_\r=-\pa_\r\log \Phi^2$,
i.e. the Weyl vector field is of ``pure gauge type" possessing zero
length curvature, $f_{\mn}=0$, determining thereby the $D(1)$-part
of the connection on the Weyl frame bundle and, in fact, reducing this
bundle completely to a bundle over a $V_4$
for a {\sl constant} modulus $\Phi$ as a consequence of the field
equations for $\ka_\r$ in the broken theory.

It is clear from the role the modulus of the scalar field plays
in this theory by acting as a potential for the Weyl vector 
field -- which itself is part of the original Weyl geometry --
that the scalar field with nonlinear selfcoupling is not a
true matter field describing scalar particles. It is a universal
field necessary to establish a scale of length in a theory
and should probably not be interpreted as a field having
a particle interpretation. This raises the question whether it
was reasonable to endow this field with a charge, denoted by $q$,
and couple it to the electromagnetic fields. The question whether this
charge has to be zero or is nonvanishing can not be decided in the
present theory. This topic will be taken up again in connection
with the study of an additional $SU(2)$ symmetry (weak isospin)
and a corresponding representation character of $\vphi$ with
respect to this group. The extension of the group structure from
$U(1)$ to $U(1)\otimes SU(2)$ will be addressed in a separate
paper devoted to an investigation of a Weyl-invariant theory
and its breaking for a case which is closest to the situation
realized in the unified electroweak theory.

The result $\Phi = const>0$ yielding a formulation of the theory
in a Riemannian space $V_4$ leads, for the fermion field $\psi$,
to the mass value $M=\Phi\,m$ relative to the scale set by the
scalar field after Weyl-symmetry breaking. The modulus $\Phi$
of the scalar field, however, also enters the expression
for the gravitational constant having a Brans-Dicke-like form
as discussed in Sect. V (compare Eq. (\ref{54}) for $a=1$).
It appears that this mass shift for the $\psi$-field cannot
yield arbitrary large values for $M$ since, at the same time, this
would reduce the gravitational coupling constant $\ka_{\hbox {\tiny E}}$
of the theory correspondingly. It was always a surprising feature of
the conventional Higgs mechanism that masses could be shifted to
arbitrary large values in the usual theory of spontaneous symmetry breaking
triggered by a nonvanishing vacuum expectation value of the scalar
field. No limitations for the size of the actual mass value
seem to arise in the usual formalism which would, in fact, be expected to
appear if gravitation were included in the theory. In the broken Weyl
theory studied in this paper gravitation is present from the outset,
and it is seen from the broken theory formulated in a $V_4$ that
the capacity of fermionic matter to generate gravitational fields
is {\sl diminished} when the value of $M$ becomes very large due to the
modulus $\Phi$ becoming very large compared to unity in Eq. (5.12). Thus
the mass giving phenomenon due to $D(1)$ symmetry breaking does have
an effect on gravity when studied in this Weyl-geometric framework
athough it is still not possible to determine the actual value of $\Phi$.

A further question regards the quantum nature of the scalar and
spinor fields involved. So far, the theory presented in this paper
is formulated in a quasi classical, single-particle description
for the $\vphi$- and $\psi$-fields with the modulus $\Phi$
of $\vphi$ being, indeed,  a classical field corresponding, as mentioned, to
the vacuum expectation value of the scalar field in the conventional
formulation of spontaneous symmetry breaking in gauge theories.
Although it does not seem to make much sense to give to the scalar
field in the present theory a fully quantized many-particle (i.e.
second quantized) form, it may nevertheless be attractive to extend the description
of the fermion matter field to a many-particle formalism by
introducing Fock space fibers in an associated bundle defined in analogy
to the bundle (2.8). It may also be interesting to
relate such a many-particle formalism to Prugove\v cki's programme
of stochastic quantization and use the arbitrary spin fields
introduced and discussed in \cite{WDD} in order to extend the
presented Weyl-geometric description to a theory involving
geometro-stochastically quantized matter fields $\Psi$ of
arbitrary spin.

\newpage
\begin{appendix}
\section
{The Geometry of a Weyl space $W_4$}

A natural basis in the local tangent space $T_x(W_4)$ at $x\in W_4$
will be denoted by $\pa_{\mu}$; $\mu $ = 0,1,2,3, and, similarly, a natural
basis in the dual tangent space $T_x^*(W_4)$ at $x\in W_4$ will be
denoted by $dx^{\mu}$ with the corresponding contravariant and
covariant tensor components labelled with Greek indices.
A local Lorentzian basis and cobasis is denoted by $e_i$ and
$\theta^i$; i = 0,1,2,3, respectively, and the corresponding tensor
components referring to an orthonormal frame are labelled by
Latin indices.
Greek indices are lowered and raised with $g_{\mn}$  and
$g^{\mn}$, respectively, while local Lorentzian indices are lowered and
raised by the metric tensor of Minkowski space $\eta_{ik}$ and $\eta^{ik}$,
respectively. The summation convention for twice appearing Greek or Latin
indices is always assumed.

Historically a connection in a $W_4$ is given in a natural basis by the
following connection coefficients:

\begin{equation}
\Gamma_{\mn}{}^{\r} = \ha g^{\r\l} \left(\pa_{\mu}g_{\nu\l}
    + \pa_{\nu}g_{\mu\l} - \pa_{\l} g_{\mn}\right)
 -\ha \left(\ka_{\mu}\d^{\r}_{\nu}+\ka_\nu\d^{\r}_{\mu}-\ka^{\r} g_{\mn}\right) =
 \bar\Gamma_{\mn}{}^{\r} + W_{\mn}{}^{\r}
\label{A1}
\end{equation}
expressed in terms of the Christoffel symbols,
$\bar\Gamma_{\mn}{}^{\r} = \{{\r \atop\mn}\}$
determined by the metric $g_{\mn}$, and by a Weyl
addition, called $W_{\mn}{}^{\r}$, determined by the Weyl
vector field $\ka_{\r}$.
Here and in the following discussion we denote purely metric
quantities pertaining to a $V_4$ by a bar.
The fourty coefficients $\Gamma_{\mn}{}^{\r}$
defined by (\ref{A1}) are symmetric in $\mu $ and $\nu $, 
i.e.~$\Gamma_{\mn}{}^\r = \Gamma_{\nu\mu}{}^\r$.
Moreover, they are invariant under Weyl transformations (1.2), (1.3)
allowing thus the definition of a covariant derivative without
specifying a particular metric and Weyl vector field in the class (1.1).

A tensor field $\phi^{(n,m)}(x)$ of type $(n,m)$, i.e. being
covariant of degree $n$ and contravariant of degree $m$, has
Weyl weight $w(\phi^{(n,m)})$ if it transforms under Weyl transformations
(1.2), (1.3) as
\begin{equation}
\phi^{(n,m)'}(x) = [\si(x)]^{w(\phi^{(n,m)})}~ \phi^{(n,m)}(x)\,.
\label{A2}
\end{equation}
The Weyl-covariant derivative of $\phi^{(n,m)}$, i.e.~the covariant
derivative $D = dx^\r D_{\r}$ invariant under (1.2), (1.3) as well as
under changes of the atlas on $W_4$, is given by (we suppress the
arguments $(x)$)
\begin{equation}
D\phi^{(n,m)} = \nabla \phi^{(n,m)} - w(\phi^{(n,m)})~\ka~\phi^{(n,m)}\,,
\label{A3}
\end{equation}
where $\nabla = dx^{\r} \nabla_\r$
with $\nabla_\r$ denoting the covariant derivative
with respect to $\Gamma_{\mn}{}^{\r}$ in the direction $\pa_{\r}$.
Clearly, $D\phi^{(n,m)}$ transforms under Weyl transformations
in the same manner as $\phi^{(n,m)}$ does.

Eq. (1.2) shows that the covariant metric tensor has Weyl weight
$w(g_{\mn}) = 1$ implying that 
\begin{equation}
D_{\r}g_{\mn} = 0 \qquad \mbox{is equivalent to} \qquad
\nabla_{\r}g_{\mn} = \kappa_\r~ g_{\mn}\,.
\label{A4}
\end{equation}

Eq. (\ref{A4}) is a Weyl-covariant statement which, for $\kappa_{\r} = 0$,
goes over into the relation $\bar\nabla_{\r}~g_{\mn} = 0$
known from Riemannian geometry with
$\bar\nabla_{\r}$
denoting the metric covariant derivative with respect to
$\bar\Gamma_{\mn}{}^{\r}$.
Analogously, the contravariant metric tensor $g^{\mn}$ has Weyl weight
$w(g^{\mn}) = -1$, with the determinant $g$ of $g_{\mn}$ having Weyl
weight $w(g) = 4$.

Let us now consider local Lorentzian frames and coframes defined by
\begin{equation}
e_i = \l_i^{\mu}(x)\pa_{\mu};\qquad \theta^i = \l^i_{\mu}(x)dx^{\mu}\,,
\label{A5}
\end{equation}
where $\l^i_\mu(x)$ and $\l^\mu_i(x)$ are the vierbein fields and
their inverse, respectively,
and characterize the geometry of a $W_4$ by structural equations of
Cartan type for the connection on the Weyl frame bundle
\begin{equation}
P_{\w}\Bigl(W_4, G = SO(3,1) \otimes D(1)\Bigr)
\label{A6}
\end{equation}
over a base $W_4$ with the structural group G being the direct product
of the orthochronous Lorentz group $SO(3,1) \equiv O(3,1)^{++}$ and the
dilatation group $D(1)$. (In order to make this
paper selfcontained we repeat here some of the formulae appearing already
in Ref. \cite{DH}.)

The bundle $P_{\w}$ can be thought of as the reduction
of the generalized linear frame bundle with structural group $Gl(4,R)$
due to the introduction of a metric with signature $(+,-,-,-)$ on the
space-time base manifold with $g_{\mn}$ taking values in the coset
space $Gl(4,R)/SO(3,1)$.
In $P_{\w}$ the metric is, however, only fixed modulo Weyl transformations.
Moreover, the connection on the general linear frame bundle reduces
to a connection on $P_{\w}$ if the metric is Weyl-covariant constant,
i.e.~satisfies (A4) (compare Ref. \cite{KN}).

We first observe that the vierbein fields $\l_{\mu}^i(x)$ have Weyl weight
$w(\l_{\mu}^i) = \ha$, while their inverse, $\l_i^{\mu}(x)$, have Weyl
weight $w(\l_i^{\mu}) = -\ha$.
The same applies to $\theta^i$ and $e_i$, respectively. As usual the
relation between metric and vierbein fields is given by
\begin{equation}
g_{\mn}(x) = \l^i_{\mu}(x) \l^k_\nu(x)~ \eta_{ik}
\label{A7}
\end{equation}
with $\eta_{ik} = {\rm diag} (1,-1,-1,-1)$ assumed to have Weyl weight zero.
The above convention adopted for the vierbein fields
and their inverse  implies that changing a Greek tensor index of a
quantity into a Latin (local Lorentzian) index changes the Weyl weight
of the components by half a unit.

The pull back of a connection on $P_{\w}$ is given by a set of one-forms
$(w_{ik}, \ka)$, obeying
\begin{equation}
\om_{ik} = -\om_{ki} = \theta^j \Gamma_{jik}; \quad \ka = \theta^j \ka_j\,,
\label{A8}
\end{equation}
where $\Gamma_{jik}$ and $\ka_j$ are the connection coefficients for the
Lorentz part and the $D(1)$ part, respectively.

Cartan's formula for the relation between the $\bar\Gamma_{\mn}{}^\r$
and the Ricci rotation coefficients
$\bar \Gamma_{jik}$ of a $V_4$ --~neglecting torsion in the context of
this paper, which was part of Cartan's original 
argument in \cite{C}~--
may now be extended to include the Weyl vector field
$\ka_i = \l_i^{\mu}\,\ka_{\mu}$
and can be written in Weyl-covariant form as
\begin{equation}
D\l_i^{\mu} \equiv \widehat{\nabla}\l_i^{\mu} + \ha\ka\l_i^{\mu}
- \omega_i{}^j \l_j^{\mu} = 0\,,
\label{A9}
\end{equation}
where $\widehat{\nabla}\l_i^{\mu}$ denotes the covariant derivative of
only the Greek index of $\l_i^{\mu}$ with respect to 
$\Gamma_{\mn}{}^{\r}$ as defined in (\ref{A1}).
Equation (\ref{A9})
is consistent with $Dg_{\mn} = 0$ and equation (\ref{A7})
as well as with the antisymmetry of the forms $\omega_{ik}$
expressed in (\ref{A8}). We mention in passing that $\ka_{\mu}$
has Weyl weight zero; correspondingly, $\ka_i$ has Weyl weight
$-\ha$, and $\ka = \ka_{\mu}dx^{\mu} = \ka_i\theta^i$ is a one-form
of Weyl weight zero.
Equation (\ref{A9}) now yields for the Lorentz part of the connection
on $P_{\w}$ the result
\begin{equation}
\omega_{ik} = \overline{\omega}_{ik} -\ha(\ka_i\theta_k-\ka_k\theta_i),
\label{A10}
\end{equation}
where $\overline{\omega}_{ik} = \theta^j \bar \Gamma_{jik}$ is the metric
part with $\bar \Gamma_{jik}=-\bar \Gamma_{jki}$
denoting the Ricci rotation coefficients
of a $V_4$. It is easy to show using (\ref{A1}) and (\ref{A9})
that the $\omega_{ik}$ are invariant under Weyl transformations
(1.2), (1.3) obeying $\omega'_{ik} = \omega_{ik}$.

Remembering that the Weyl weight of $\theta^k$ is $\ha$ the structural
equations of Cartan type characterizing the geometry of a $W_4$ may
now be written in Weyl-covariant form as:
\begin{eqnarray}
D\theta^k \equiv d\theta^k+\omega_j{}^k\wedge\theta^j-
\ha\ka\wedge\theta^k &=& 0\,,\\
\label{A11}
d\omega_{ik} + \omega_{ij}\wedge\omega_k{}^j &=& \Omega_{ik}\,,\\
\label{A12}
d\ka &=& f\,.
\label{A13}
\end{eqnarray}
Here Eq. (A11)
states that the torsion vanishes in a $W_4$, and
Eqs. (A12) and (\ref{A13})
define the curvature two-forms ($\Omega_{ik}, f$) on
$P_{\w}$ (pulled back to the base) of the connection one-forms
$(\om_{ik}, \ka)$ given by
(\ref{A8}) and (\ref{A10})
with $\Omega_{ik} = -\Omega_{ki}$.

The Bianchi identities for a $W_4$ follow in the usual way
as integrability conditions from the structural equations
(A11) -- (A13)
by exterior derivation. They read:
\begin{equation}
[~\Omega_{jk} - \ha f\eta_{jk}]\wedge\theta^j = 0\,,
\label{A14}
\end{equation}
\begin{equation}
D\Omega_{ij} = d\Omega_{ij} - \omega_i{}^k\wedge\Omega_{kj} -
\omega_j{}^k\wedge\Omega_{ik}  = 0\,.
\label{A15}
\end{equation}

If torsion is nonzero, i.e.~for a Cartan-Weyl space $CW_4$
(compare Ref.\cite{DH}) the rhs of
Eq. (\ref{A14}) would be given by $D\Omega_k$,
where $\Omega_k$ is the vector valued torsion two-form.

In parenthesis we would like to remark that one could also introduce for
the characterization of a $W_4$ a connection form
\begin{equation}
\widehat{\om}_{ik} = \om_{ik} -\ha\ka\eta_{ik}
\label{A16}
\end{equation}
which is no longer Lorentz Lie algebra valued (i.e.~antisymmetric
in $i,k$) and write Eq. (A11) as
\begin{equation}
D \theta^k \equiv d \theta^k + \widehat{\omega}_j{}^k\wedge\theta^j = 0\,.
\label{A17}
\end{equation}
Rewriting the structural equations (A12) and (A13) in terms
of $\widehat{\omega}_{ik}$ yields
\begin{equation}
d\widehat{\omega}_{ik} + \widehat{\omega}_{ij}\wedge\widehat{\omega}_k{}^j =
\widehat{\Omega}_{ik} = \Omega_{ik} -\ha f\eta_{ik}\,.
\label{A18}
\end{equation}
Here $\widehat{\Omega}_{ik}$
represents the total Weyl curvature, which appears already in
(\ref{A14}), being composed of Lorentz curvature,
$\Omega_{ik} = -\Omega_{ki}$, and $D(1)$ or ``length curvature''
$f$. It is often useful to introduce a curvature tensor which is
not antisymmetric in the second pair of indices being thus of the
type defined on the rhs of Eq. (\ref{A18}). We shall do  this below
and in the text for the Greek indexed curvature tensor of a $W_4$
which is ``of $~\widehat {}~$ type'' although we omit the ``hat'' for
simplicity in denoting the tensor components.

The $W_4$-curvature tensor is obtained from the two-forms
(A12) by an expansion in a basis of two-forms:
\begin{equation}
\Omega_{ik} = \ha \theta^j \wedge \theta^l R_{jlik}\,.
\label{A19}
\end{equation}
In the following we prefer to work mainly with the tensor
$R_{\mn\r\l}$ defined by
\begin{equation}
\widehat\Omega_{ik}\l_\r^i \l_\l^k = \ha dx^{\mu}\wedge dx^\nu R_{\mn\r\l},
\label{A20}
\end{equation}
where we suppress, as mentioned, for simplicity the hat on
$R_{\mn\r\l}$.

It is easy to show that the tensor 
$R_{\mn\r}{}^{\si} = R_{\mn\r\l}g^{\l\si}$
is invariant under Weyl transformations obeying
\begin {equation}
R_{\mn\r}'{}^{\si} = R_{\mn\r}{}^{\si}\,.
\label{A21}
\end{equation}
The curvature tensor $R_{\mn\r}{}^\si$ of a $W_4$ is composed of a
metric part, $\bar R_{\mn\r}{}^\si$, and a Weyl addition denoted
by $P_{\mn\r}{}^\si$ which is expressed in terms of $\ka_\r$ and its
metric covariant derivatives (see below). Correspondigly, the
contracted tensor
\begin {equation}
R_{\mu\r} = R_{\mn\r}{}^\nu = \bar R_{\mu\r} + P_{\mu\r}
\label{A22}
\end{equation}
is Weyl-invariant and decomposes into a metric part
(the Ricci tensor) and a Weyl addition called $P_{\mu\r}$.
The curvature scalar
\begin {equation}
R = R_{\mn}g{}^{\mu\nu} = \bar R + P
\label{A23}
\end{equation}
is a scalar of Weyl weight $w(R) = -1$.

Explicitly, the tensor $R_{\mn\r\si}$ with Weyl weight 1
may be split into
\begin {equation}
R_{\mn\r\si} = \bar R_{\mn\r\si} + P_{\mn\r\si}
\label{A24}
\end{equation}
with the Riemann-Christoffel tensor given by
$(\bar \Gamma_{\mn\si} = \bar \Gamma_{\mn}{}^\r g_{\r\si})$:
\begin {equation}
\bar R_{\mn\r\si} = \pa_\mu \bar \Gamma_{\nu\r\si}-
                    \pa_\nu \bar \Gamma_{\mu\r\si}+
\bar \Gamma_{\mu\r\l} \bar \Gamma_{\nu\si}{}^\l -
\bar \Gamma_{\nu\r\l} \bar \Gamma_{\mu\si}{}^\l\,,
\label{A25}
\end{equation}
and with the Weyl addition
\begin{eqnarray}
P_{\mn\r\si} = -\ha(g_{\mu\r} \bar \nabla_\nu \ka_\si +
                   g_{\nu\si} \bar \nabla_\mu \ka_\r -
                   g_{\mu\si} \bar \nabla_\nu \ka_\r - 
                   g_{\nu\r} \bar \nabla_\mu\ka_\si)\nonumber\\
\qquad   -\frac{1}{4}(g_{\mu\r} \ka_\nu \ka_\si +
³                  g_{\nu\si} \ka_\mu \ka_\r - g_{\mu\si}\ka_\nu\ka_\r
                  -g_{\nu\r} \ka_\mu \ka_\si)\nonumber\\
\qquad   +\frac{1}{4}(g_{\mu\r} g_{\nu\si} - g_{\mu\si} g_{\nu\r})
                   \ka_\l \ka^\l - \ha f_{\mn} g_{\r\si}.
\label{A26}
\end{eqnarray}

Clearly, the splitting in (\ref{A24}) and in the contractions (\ref{A22})
and (\ref{A23}) is not Weyl-invariant. Only the sums of the two terms
on the rhs of these equations has a definite covariant behaviour
under Weyl transformations reproducing the tensor except for a factor
$[\si(x)]^w$ with the mentioned weight $w$. When we consider a splitting
of a quantity into a metric (i.e. $V_4$) part and a Weyl addition it is
always understood that this split is made in a particular Weyl gauge and
that it is in general not invariant against Weyl transformations.

Besides the Ricci tensor $\bar R_{\mn} $, being symmetric in $\mu, \nu$, the
Weyl addition in (\ref{A22}) is, from (\ref{A26}), given by
\begin{equation}
P_{\mn} = P_{(\mn)} + P_{[\mn]}
\label{A27}
\end{equation}
with
\begin{equation}
P_{(\mn)} = \ha(P_{\mn} + P_{\nu\mu}) =
           -\ha(\bar \nabla_\mu \ka_\nu+\bar \nabla_\nu \ka_\mu)
           -\ha g_{\mn} \bar \nabla^\r \ka_\r + \ha g_{\mn} \ka^\r \ka_\r
           -\ha \ka_\mu \ka_\nu\,,
\label{A28}
\end{equation}
and
\begin{equation}
P_{[\mn]} = \ha(P_{\mn}-P_{\nu\mu}) \equiv R_{[\mn]} = - f_{\mn}\,,
\label{A29}
\end{equation}
where the length curvature tensor is defined by
\begin{equation}
f_{\mn} = \bar \nabla_\mu \ka_\nu - \bar \nabla_\nu \ka_\mu =
        \pa_\mu \ka_\nu - \pa_\nu \ka_\mu\,.
\label{A30}
\end{equation}
The curvature scalar (\ref{A23}) of a $W_4$ is, finally, given by
\begin{equation}
R = \bar R -3 \bar \nabla^\r \ka_\r +\frac{3}{2} \ka^\r \ka_\r\,.
\label{A31}
\end{equation}

The tensor $R_{\mn\r\si}$ defined by (\ref{A20}) is antisymmetric in
its first two indices and has symmetric and antisymmetric
contributions regarding the last two indices.
Below we shall derive from it a curvature
tensor (the Bach tensor) having the same symmetries as the 
Riemann-Christoffel tensor in a $V_4$. Contracting the last two
indices in (\ref{A24}) with $g^{\r\si}$ yields
\begin{equation}
R_{\mn\r\si} g^{\r\si} = P_{\mn\r\si} g^{\r\si} = - 2f_{\mn}\,.
\label{A32}
\end{equation}

Sometimes the following formula is useful showing the lack of
antisymmetry in the last two indices of the full $W_4$ curvature tensor:
\begin{equation}
R_{\mn\r\si} = - R_{\mn\si\r} - f_{\mn} g_{\si\r}\,.
\label{A33}
\end{equation}
This formula together with (A22) and (A29) implies that
$R^\r{}_{\mu\r\nu}=R_{\mn}+f_{\mn}=R_{(\mn)}$\,.

The commutator of two Weyl-covariant derivatives of a contravariant
vector $a^\r$ with Weyl weight $w(a^\r)$ is given by
\begin{equation}
\left[D_\mu, D_\nu \right]a^\r = R_{\mn\si}{}^\r a^\si
  - w(a^\r)f_{\mn} a^\r\,.
\label{A34}
\end{equation}
With the help of (\ref{A33}) and the relation $w(a_\r) = w(a^\r)+1$
one immediately derives the corresponding relation for
$a_\r = g_{\r\si} a^\si$:
\begin{equation}
\left[D_\mu, D_\nu \right]a_\r = -R_{\mn\r}{}^\si a_\si
   -w(a_\r)f_{\mn} a_\r\,.
\label{A35}
\end{equation}
Analogous formulae are obtained for tensors $a^{\mn\dots}_{\r\l\dots}$
of higher rank with a curvature term appearing on the rhs for each
contravariant or covariant index analogous to Eqs. (\ref{A34})
and (\ref{A35}), respectively.

The cyclic conditions following from (\ref{A14}) for the curvature
tensor $R_{\mn\r\si}$ read
\begin{equation}
R_{\{\mn\r\}\si} = 0\,,
\label{A36}
\end{equation}
with $\{\mn\r\}$ denoting the cyclic sum of the indices in the curly
brackets. Finally, the Bianchi identities for the curvature tensors
$R_{\mn\r\si}$ and $f_{\mn}$
[compare (\ref{A15}) and (\ref{A13})] read:
\begin{eqnarray}
D_{\{\l} R_{\mn\}\r\si} &=& 0\,,\\
\label{A37}
D_{\{\l} f_{\mn\}} = \pa_{\{\l} f_{\mn\}} &=& 0\,.
\label{A38}
\end{eqnarray}
Contraction of (A37) with $g^{\nu\si}$ yields the Ricci-type
identities for a $W_4$
\begin{equation}
D^\r R_{\mn\si\r} = - (D_\mu R_{\nu\si}-D_\nu R_{\mu\si})\,.
\label{A39}
\end{equation}
A second contraction, finally, leads to a formula which we would like to
quote in a form used in the text:
\begin{equation}
D^{\nu}\left(R_{(\mn)}-\ha g_{\mn}R\right) = \ha D^{\nu} f_{\mn}\,.
\label{A40}
\end{equation}
The factor $\ha $ on the rhs of (\ref{A40}) is due to the fact
that this equation derives from (\ref{A15}) and (\ref{A19}) and
not from (\ref{A20}) involving the curvature two-form
$\widehat \Omega_{ik}$, although we have used the contracted full
curvature $R_{\mn}$ to formulate the result.

Bach \cite{B}
has introduced a $W_4$ curvature tensor possessing the same symmetries
as the Riemann-Christoffel tensor of a $V_4$:
\begin{equation}
S_{\mn\r\si} = S_{([\mn][\r\si])} = \frac{1}{4}
\left\{R_{\mn\r\si}-R_{\mn\si\r}+R_{\r\si\mn}-R_{\r\si\nu\mu}\right\}\,.
\label{A41}
\end{equation}
Here we have exhibited the symmetry of the tensor by square and round
brackets. Using (\ref{A24}) and (\ref{A26}) yields the following form
for this tensor:
\begin{equation}
S_{\mn\r\si} = R_{\mn\r\si} + \frac{1}{4}
   \left(g_{\mu\r} f_{\nu\si}+g_{\nu\si} f_{\mu\r}-g_{\mu\si}
  f_{\nu\r}-g_{\nu\r} f_{\mu\si}\right) + \ha f_{\mn} g_{\r\si}\,.
\label{A42}
\end{equation}
It is easy to show using (\ref{A36}) that it, moreover, satisfies the
cyclic conditions
\begin{equation}
S_{\{\mn\r\}\si} = 0\,.
\label{A43}
\end{equation}

The first contraction of (\ref{A42}) yields the symmetric tensor
of Weyl weight zero [compare Eqs. (\ref{A22}) and 
(\ref{A27}) -- (\ref{A29})]:
\begin{equation}
S_{\mu\r}=S_{\mn\r\si} g^{\nu\si} = R_{\mu\r}+f_{\mu\r} = R_{(\mu\r)}
= \bar R_{\mu\r}+P_{(\mu\r)}\,,
\label{A43a}
\end{equation}
leading to
\begin{equation}
S = S_{\mu\r} g^{\mu\r} = R\,,
\label{A43b}
\end{equation}
with $R$ as given by (\ref{A31}).

We finally write down the Weyl tensor, $C_{\mn\r\si}$,
for a $W_4$ characterized by
the property of having vanishing
contraction i.e. $C_{\mn\r}{}^{\nu}=0$:
\begin{equation}
C_{\mn\r\si} = R_{\mn\r\si}-\ha(g_{\mu\r}R_{\nu\si}+g_{\nu\si}
   R_{\mu\r}-g_{\mu\si}R_{\nu\r}-g_{\nu\r}R_{\mu\si})
   +\frac{1}{6}(g_{\mu\r}g_{\nu\si}-g_{\mu\si}g_{\nu\r})R\,.
\label{A44}
\end{equation}
This tensor may again be decomposed into a purely metric part,
$\bar C_{\mn\r\si}$, and a $W_4$ addition which we call
$K_{\mn\r\si}$:
\begin{equation}
C_{\mn\r\si} = \bar C_{\mn\r\si} + K_{\mn\r\si}\,,
\label{A45}
\end{equation}
with the Weyl tensor of a $V_4$ as given by the usual expression
\begin{equation}
\bar C_{\mn\r\si} = \bar R_{\mn\r\si}-\ha
   (g_{\mu\r}\bar R_{\nu\si}+g_{\nu\si}\bar R_{\mu\r}-g_{\mu\si}
   \bar R_{\nu\r}-g_{\nu\r}\bar R_{\mu\si})
   +\frac{1}{6}(g_{\mu\r}g_{\nu\si}-g_{\mu\si}g_{\nu\r})\bar R\,,
\label{A46}
\end{equation}
and the additional tensor $K_{\mn\r\si}$ which is expressible
in terms of the $f_{\mn}$ only:
\begin{equation}
K_{\mn\r\si} = \frac{1}{4}(g_{\mu\r}f_{\nu\si}+g_{\nu\si}f_{\mu\r}
   -g_{\mu\si}f_{\nu\r}-g_{\nu\r}f_{\mu\si})
   -\ha f_{\mn} g_{\r\si}\,.
\label{A47}
\end{equation}

On the other hand, expressing on the rhs of (A46) the
curvature tensor $R_{\mn\r\si}$ and its contractions by the
Bach tensor (\ref{A42}) and its contractions (\ref{A43a}) and
(\ref{A43b}) one at once derives the following result:
\begin{equation}
C_{\mn\r\si} = S_{\mn\r\si} - \ha (g_{\mu\r}S_{\nu\si} +
g_{\nu\si}S_{\mu\r} - g_{\mu\si}S_{\nu\r} - g_{\nu\r}S_{\mu\si}) +
\frac{1}{6}(g_{\mu\r}g_{\nu\si}
- g_{\mu\si}g_{\nu\r}) S
+ K_{\mn\r\si}\,.
\label{A47a}
\end{equation}
This shows in comparing with (\ref{A45}) that the first three terms
on the rhs of this equation define a curvature tensor (possessing
vanishing contraction) which is {\sl independent} of $\ka_\r$
yielding thus
\begin{equation}
S_{\mn\r\si} - \ha (g_{\mu\r}S_{\nu\si} + g_{\nu\si}S_{\mu\r} -
g_{\mu\si}S_{\nu\r} - g_{\nu\r}S_{\mu\si}) +
 \frac{1}{6}(g_{\mu\r}g_{\nu\si} -
g_{\mu\si}g_{\nu\r}) S = \bar C_{\mn\r\si}\,.
\label{A47b}
\end{equation}

It is well known that the tensor $\bar C_{\mn \r}{}^{\si}$
is conformally invariant. Hence $K_{\mn\r}{}^\si$ in (\ref{A45}) is
Weyl-invariant by itself since $C_{\mn\r}{}^\si$ is Weyl-invariant
(having Weyl weight zero). Therefore the splitting of the rhs
of (\ref{A45}), when the index $\si$ is raised in this equation,
is indeed Weyl-invariant.

Furthermore, one has
\begin{equation}
C_{\mn\r\si}g^{\r\si} = C_{\mn\r}{}^\r = K_{\mn\r}{}^\r = -2f_{\mn}\,.
\label{A48}
\end{equation}

The comparison of the right-hand sides of Eqs. (\ref{A42}) and
(\ref{A47}), finally, shows that the tensor $S_{\mn\r\si}$
possessing the symmetries indicated in Eqs. (\ref{A41}) and
(\ref{A43}) can be expressed as
\begin{equation}
S_{\mn\r\si} = R_{\mn\r\si}+K_{\mn\r\si}+f_{\mn}g_{\r\si}
\label{A49}
\end{equation}
with an analogous decomposition of the tensor $S_{\mn\si}{}^\r$
of Weyl weight zero being Weyl-invariant.

We close this appendix by mentioning the curvature invariant
of Weyl weight $-2$ constructed with the help of the tensor
$C_{\mn\r\si}$ (compare Bach \cite{B}). Using (\ref{A45}) and
(\ref{A47}) one finds
\begin{equation}
C_{\mn\r\si}C^{\mn\r\si} = \bar C_{\mn\r\si}\bar C^{\mn\r\si}
+ \frac{3}{2} f_{\mn}f^{\mn}\,,
\label{A50}
\end{equation}
where the first term on the rhs may be rewritten in terms of
$S_{\mn\r\si}$, $S_{\mn}$, and $S$ using (A51).
When multiplied with $\sqrt{-g}$ Eq. (\ref{A50}) yields a
quadratic curvature invariant of Weyl weight zero which may
be considered as a possible Lagrangean density for a theory
based on a $W_4$. The rhs of (\ref{A50}), moreover, shows
that this invariant is composed of an invariant expression
{\sl independent} of $\ka_\r$ and an invariant constructed
from the $D(1)$-curvature $f_{\mn}$ alone.

\section{}
For completeness we collect in this appendix some (partly well-known)
formulae used in the text involving the variation of the
square root of the determinant
of the metric tensor, $g_{\mn}$, having Weyl weight $w(\sqrt{-g})= -2$,
and the variation of the vierbein fields under metric variations.

Varying the metric tensor one obtains the following relations:
\begin{eqnarray}
\delta\sqrt{-g} &=& -\ha\sqrt{-g}~g_{\mn} \delta g^{\mn},\\
\label{B1}
g_{\mn} \delta g^{\mn} &=& -g^{\mn} \delta g_{\mn}\,.
\label{B2}
\end{eqnarray}

The variations of the vierbein fields are related to the variation
of the metric by
\begin{equation}
\delta \l_\mu^k = -\ha \l_\si^k~ g_{\mu\r}~ \delta g^{\r\si}; \qquad
\delta \l_j^\nu = +\ha \l_j^\mu~ g_{\mu\r}~ \delta g^{\r\nu}.
\label{B3}
\end{equation}
This implies, for example, for the metric variation of a gradient,
$\pa_k\vphi$, of a function $\vphi$ expressed in a local
Lorentzian frame the relation
\begin{equation}
\delta\pa_k\vphi = \ha \eta_{ki} \l_\r^i \l_\mu^j \delta g^{\r\mu}
\pa_j\vphi\,.
\label{B4}
\end{equation}
Moreover, one finds from (\ref{A1}) the formulae:
\begin{eqnarray}
\Gamma_{\mu\r}{}^\r &=& \pa_{\mu} \log \sqrt{-g}~ -~ 2\ka_{\mu}\,,\\
\label{B5}
g^{\mn} \Gamma_{\mn}{}^\r &=& - \frac{1}{\sqrt{-g}} \pa_{\mu}
   (\sqrt{-g}~g^{\mu\r}) ~+~ \ka^\r\,.
\label{B6}
\end{eqnarray}

\section{The Euler-Gauss-Bonnet and Pontrjagin Invariants for a $W_4$}

There is a particular combination with Weyl weight $-2$ composed of
quadratic curvature invariants $S_{\mn\r\si}S^{\mn\r\si}, S_{\mn}S^{\mn}$,
and $S^2$ which yields, when multiplied with $\sqrt{-g}$, an invariant
density (a scalar of Weyl weight zero) which plays a particular role in
the search for a possible Lagrangean density in a variational formulation
of a theory based on a Weyl geometry. Bach \cite{B} found that the
following combination of quadratic curvature invariants leads to a
contribution in a variational derivation of the field equations which
is identically zero:
\begin{eqnarray}
\sqrt {-g}~\{ S_{\mn\r\si}S^{\mn\r\si} - 4 S_{\mn}S^{\mn} &+& S^2
                 + \ha f_{\mn}f^{\mn} \} \nonumber\\
 &=& \sqrt {-g}~\{ R_{\mn\r\si}R^{\mn\r\si} -
             4 R_{\mn}R^{\mn} + R^2 + 3f_{\mn}f^{\mn}\}\,.
\label{C1}
\end{eqnarray}

To represent (\ref{C1}) in a form making reference to the analogue of the
Euler-Gauss-Bonnet invariant of a $V_4$ (see Chern \cite{Ch}) for the
$W_4$ case discussed here, one considers the four-form
\begin{equation}
{\cal E} = \frac{1}{32\pi^2}~\varepsilon_{ijkl}~\Omega^{ij}\wedge
           \Omega^{kl} = -\frac{1}{32\pi^2}~I\eta\,,
\label{C2}
\end{equation}
where $\Omega^{ij}$ is the curvature two-form defined in 
(A12), and $\eta = \theta^0\wedge\theta^1\wedge\theta^2
\wedge\theta^3=\sqrt{-g}d^4x$ is the volume form.
The invariant $I$ in
(\ref{C2}) is given by (compare (\ref{A19}); see also
Ref. \cite{WD} for the case of a Riemann-Cartan space
$U_4$):
\begin{equation}
I = -\frac{1}{4} \varepsilon_{rspq}~
\varepsilon_{ijkl}~R^{rsij}~R^{pqkl}
 = R_{ijkl}R^{klij} - 4 R_{kj}R^{jk} + R^2 \,.
\label{C3}
\end{equation}
One finds that the curly brackets in Eq. (\ref{C1}) can be 
rewritten as:
\begin{equation}
S_{\mn\r\si}S^{\mn\r\si} - 4 S_{\mn}S^{\mn} + S^2 + \ha f_{\mn}f^{\mn}
= \underline{R}_{\mn\r\si}\underline{R}^{\r\si\mn} -
4 \underline{R}_{\mn}\underline{R}^{\nu\mu}
       + \underline{R}^2 \,,
\label{C4}
\end{equation}
where we used the notation
\begin{equation}
\underline{R}_{\mn\r\si}=R_{\mn\r\si}+\ha f_{\mn}g_{\r\si}\,;\quad
\underline{R}_{\mn}=R_{\mn}+\ha f_{\mn}\,;\quad
\underline{R}=R\,,
\label{C5}
\end{equation}
relating the Greek indexed curvature tensors associated with
$\Omega^{ik}$ and $\widehat\Omega^{ik}$, respectively, and its 
contractions [compare (\ref{A18}); to avoid confusion we use an
underline to denote the curvature tensor $\underline R_{\mn\r\si}=
\l^i_\mu\l^j_\nu\l^k_\r\l^l_\si~R_{ijkl}$ which is antisymmetric in
both pairs of indices, $\mn$ {\sl and} $\r\si$].

Thus Bach's observation that the lhs of (\ref{C1}) yields
identically zero when used under a variational integral implies that
the form ${\cal E}$ with the invariant $I$ given by (\ref{C3})
is exact, yielding upon integration in the case of a closed 
orientable (properly) Riemannian manifold the 
Euler-Poincar\'e characteristic of the manifold, and
yielding for a Riemannian submanifold with boundary the
Gauss-Bonnet theorem. As shown above, it is the
Weyl-invariant density (\ref{C1}) involving the
curvature invariant $I$ with the particular summation
over the indices shown in (\ref{C3}) and
(\ref{C4}) which represents, according to Bach's result, 
the quantity corresponding to the Euler-Gauss-Bonnet
invariant for a Weyl space $W_4$ being, moreover,
independent of the particular Weyl gauge adopted
in the family (\ref{11}).

Bach also investigated two further quadratic curvature
invariants involving the dual curvature tensors of Weyl
weights $w({}^*f_{\mn})=0$ and $w({}^*R_{\mn\r\si})=1$ :
\begin{equation}
{}^*f_{\mn} = \ha\sqrt{-g}~\varepsilon_{\mn\r\l}f^{\r\l}\,,
\quad\mbox{and}\quad
{}^*R_{\mn\r\si} = \ha\sqrt{-g}~\varepsilon_{\mn\d\l}
R^{\d\l}{}_{\r\si}\,.
\label{C6}
\end{equation}
He showed that the following invariants yield again a vanishing
contribution in a variational derivation of field equations in
a $W_4$ (compare \cite {B}, Sect. XII):
\begin{equation}
\sqrt{-g}~f_{\mn}{}^*f^{\mn}\,,\quad\mbox{and}\quad
\sqrt{-g}~R_{\mn\r\si}{}^*R^{\mn\r\si}\,.
\label{C7}
\end{equation}
The expressions (\ref{C7}) may be related to the Pontrjagin
invariant for a $W_4$ which, in analogy to the Riemannian case,
may be defined by the four-form of Weyl weight zero:
\begin{equation}
{\cal P} = \frac{1}{8\pi^2}~\Omega_{ij}\wedge\Omega^{ij} = J\,\eta\,,
\label{C8}
\end{equation}
with
\begin{equation}
J = \frac{1}{16\pi^2}~R_{ijkl}~{}^*R^{ijkl} =
\frac{1}{16\pi^2}~\underline R_{\mn\r\si}~{}^*\underline R^{\mn\r\si}\,,
\label{C9}
\end{equation}
where in the last equality in (\ref{C9}) we have used the notation
introduced in (\ref{C4}). Bach's observation that also the invariants 
(\ref{C7}) yield identically zero when used in a variational
derivation of the field equations thus implies, together with the
identity
\begin{equation}
\sqrt{-g}~\underline R_{\mn\r\si}{}^*\underline R^{\mn\r\si} =
\sqrt{-g}~(R_{\mn\r\si}{}^*R^{\mn\r\si} - f_{\mn}{}^*f^{\mn})\,,
\label{C10}
\end{equation}
and Eq. (\ref{C9}), that ${\cal P}$ is an exact form on $W_4$
which is invariant under Weyl transformations.
\end{appendix}


\begin{thebibliography}{99}
\bibitem{W} H. Weyl, {\sl Gravitation und Elektrizit\"at},
            Sitzungsber.\ Preuss.\ Akad.\ Wiss.\ 465-480 (1918); and {\sl Eine
            neue Erweiterung der Relativit\"atstheorie}, Ann.\ Physik (Leipzig)
            {\bf 59}, 101-133 (1919).
\bibitem{RZM} H. Weyl, {\sl Raum-Zeit-Materie}, 7th Edition,
            Springer-Verlag, Heidelberg 1988. Compare also: {\sl Elektron und
            Gravitation. I.}, Z. Phys. {\bf 56}, 330-352 (1929).
\bibitem{FRW} T.\ Fulton, F.\ Rohrlich and L.~Witten, {\sl Conformal
            Invariance in Physics}, Rev.\ Mod.\ Phys.\ {\bf 34}, 442-457 (1962).
\bibitem{HT} H. Tann, {\it Einbettung der Quantentheorie eines
            Skalarfeldes in eine Weyl-Geometrie --~Weyl-Symmetrie
            und ihre Brechung~--}~, Dissertation, Fakult\"at f\"ur
            Physik der Ludwig-Maximilians-Universit\"at M\"unchen, \ April 1997.
\bibitem{CCJ} C.G.\ Callan, S.\ Coleman and R.\ Jackiw, {\it A New
            Improved Energy-Momentum Tensor}, Ann.\ Phys.\ (NY) {\bf 59}, 42-73 (1970).
\bibitem{T} A. Trautman, private communication.
            See also M.\ Cahen, S.\ Gutt
             and A.\ Trautman, {\sl Spin structures on real projective quadrics},
             Journ.\ Geom.\ Phys.\ {\bf 10}, 127-154 (1993).
\bibitem{RP} R. Penrose, {\sl Zero rest-mass fields including gravitation:
             asymptotic behaviour}, Proc.\ Roy.\ Soc.\ (London) {\bf 284}, 159-203
             (1965).
\bibitem{P} W. Pauli, {\sl \"Uber die Invarianz der Dirac'schen
            Wellengleichungen gegen\"uber \"Ahnlichkeitstransformationen des
            Linienelementes im Falle verschwindender Ruhemasse},
            Helv. Phys. Acta. {\bf 13}, 204-208 (1940).
\bibitem{S} E. Schr\"odinger, {\sl Diracsches Elektron im Schwerefeld I},
            Sitzungsber.\ Preuss.\ Akad.\ Wiss. 105-128 (1932).
\bibitem{R} F. Rohrlich, {\sl Classical Charged Particles},
            Addison-Wesley Publ. Co., Redwood City 1965.
\bibitem{WDD} W. Drechsler, {\sl Geometro-stochastically quantized
             fields with internal spin variables}, J. Math. Phys.
             {\bf 38}, 5531-5558 (1997).
\bibitem{DH} W. Drechsler and D. Hartley, {\sl The role of the internal
            metric in generalized Kaluza-Klein theories},
            J. Math.Phys. {\bf 35}, 3571-3586 (1994).
\bibitem{KN} S. Kobayashi and K. Nomizu, {\sl Foundations of Differential
            Geometry}, Vol. 1, pp. 57 and 88, Interscience Publishers,
            John Wiley, New York 1963.
\bibitem{C} E. Cartan, {\sl Sur les vari\'et\'es a connexion affine et la
            th\'eorie de la relativit\'e g\'en\'eralis\'ee},
            Ann. Ec. Norm. Sup. {\bf 39}, 325-412 (1922); {\bf 41}, 1-25
            (1924), and {\bf 42}, 17-88 (1925).
\bibitem{B} R. Bach, {\sl Zur Weylschen Relativit\"atstheorie und der 
            Weylschen Erweiterung des Kr\"ummungsbegriffs}, Math. Z. {\bf 9},
            110-135 (1921).
\bibitem{Ch} S. S. Chern, {\sl A simple intrinsic proof of the
             Gauss-Bonnet formula for closed Riemannian manifolds},
             Ann. Math. {\bf 45}, 747-752 (1944), and {\sl On the
             curvatura integra in a Riemannian manifold}, Ann. Math.
             {\bf 46}, 674-684 (1945).
\bibitem{WD} W. Drechsler, {\sl Topological Invariants and the Dynamics
             of an Axial Vector Torsion Field}, GRG {\bf 15}, 703-723
             (1983).
\bibitem{H} S. Helgason, {\sl The Radon Transform}, Birkh\"auser, Basel
             1980, p. 144.
\end{thebibliography}
\end{document}